\documentclass[journal]{rmaa}

\usepackage{paralist}

\usepackage{psfrag,color}

\usepackage[latin1]{inputenc}

\usepackage{graphicx}
\usepackage{lscape}
\usepackage{txfonts}
\usepackage{natbib}
\usepackage{url}
\usepackage[T1]{fontenc}

\usepackage[bookmarks=false]{hyperref}

\newcommand{\funits}{10$^{-16}$ erg~s$^{-1}$~cm$^{-2}$~\AA$^{-1}$}

\DeclareRobustCommand{\ion}[2]{%
\relax\ifmmode
\ifx\testbx\f@series
{\mathbf{#1\,\mathsc{#2}}}\else
{\mathrm{#1\,\mathsc{#2}}}\fi
\else\textup{#1\,{\mdseries\textsc{#2}}}%
\fi}

\title{Pipe3D, a pipeline to analyze Integral Field Spectroscopy Data: I. New fitting philosophy of FIT3D.}
\author{
S.\,F. S\'anchez\altaffilmark{1},
E.\,P\'erez\altaffilmark{2},
P.\,S\'anchez-Bl\'azquez\altaffilmark{3},
J.J.\,Gonz\'alez\altaffilmark{1},
F.F.\,Rosales-Ortega\altaffilmark{4},
M.\,Cano-D\'\i az\altaffilmark{1},
C.\,L\'opez-Cob\'a\altaffilmark{1},
R.\,A. Marino\altaffilmark{5},
A. Gil de Paz\altaffilmark{5},
M. Moll\'a\altaffilmark{6},
A. R. L\'{o}pez-S\'{a}nchez\altaffilmark{7},
Y. Ascasibar\altaffilmark{8},
J. Barrera-Ballesteros\altaffilmark{9}
}

\altaffiltext{1}{Instituto de Astronom\'\i a, Universidad Nacional Auton\'oma de Mexico.}
\altaffiltext{2}{Instituto de Astrof\'{\i}sica de Andaluc\'{\i}a (CSIC), Glorieta de la Astronom\'\i a s/n, Aptdo. 3004, E18080-Granada, Spain.}
\altaffiltext{3}{Departamento de F\'isica Te\'orica, Universidad Aut\'onoma de Madrid, 28049 Madrid, Spain.}
\altaffiltext{4}{Instituto Nacional de Astrof{\'i}sica, {\'O}ptica y Electr{\'o}nica, Luis E. Erro 1, 72840 Tonantzintla, Puebla, Mexico. }
\altaffiltext{5}{CEI Campus Moncloa, UCM-UPM, Departamento de Astrof\'{i}sica y CC$.$ de la Atm\'{o}sfera, Facultad de CC$.$ F\'{i}sicas, Universidad Complutense de Madrid, Avda.\,Complutense s/n, 28040 Madrid, Spain.}
\altaffiltext{6}{Departamento de Investigaci\'on B\'asica, CIEMAT, Avda. Complutense 40 E-28040 Madrid, Spain.}
\altaffiltext{7}{Australian Astronomical Observatory, PO BOX 296, Epping, NSW 1710, Australia.}
\altaffiltext{8}{Departamento de F\'isica Te\'orica, Universidad Aut\'onoma de Madrid, 28049 Madrid, Spain.}
\altaffiltext{9}{Instituto de Astrof\'\i sica de Canarias (IAC), E-38205 La Laguna, Tenerife, Spain}

\shortauthor{S\'anchez}
\shorttitle{Pipe3D, new fitting with FIT3D}

\fulladdresses{

\item Instituto de Astronom\'\i a,Universidad Nacional Auton\'oma de Mexico, A.P. 70-264, 04510, M\'exico,D.F. (sfsanchez@astro.unam.mx)

}

\listofauthors{S. F. S\'anchez}
\indexauthor{S. F. S\'anchez}

\abstract{We present an improved version of {\sc FIT3D}, a fitting tool for 
the analysis of the spectroscopic properties of  the stellar populations
and the ionized gas derived from moderate resolution spectra of galaxies.
This tool was developed to analyze Integral Field Spectroscopy
data and it is the basis of {\sc Pipe3D}, a pipeline used in the
analysis of CALIFA, MaNGA, and SAMI data.

We describe the philosophy and each step of the fitting procedure.  We
present an extensive set of simulations in order to estimate the
precision and accuracy of the derived parameters for the stellar
populations and the ionized gas. We report on the results of those
simulations.

Finally, we compare the results of the analysis using {\sc FIT3D} with
those provided by other widely used packages, and we find that the
parameters derived by {\sc FIT3D} are fully compatible with those
derived using these other tools.

}

\resumen{Presentamos una versi\'on mejorada de {\sc FIT3D}, una herramienta de ajuste para el an\'alisis de las poblaciones estelares y el gas ionizado en espectros de galaxias de resoluci\'on intermedia. La misma se desarroll\'o para el an\'alisis de datos de espectroscop\'\i a de campo integral y es la base de {\sc Pipe3D}, un dataducto usado en el an\'alisis de datos de los muestreos CALIFA, MaNGA y SAMI.

Describimos la filosof\'\i a y los pasos seguidos en el ajuste, presentando un conjunto amplio de simulaciones con el fin de estimar la precisi\'on de los par\'ametros derivados, mostrando el resultado de dichas simulaciones.

Finalmente, comparamos el resultado del an\'alisis con {\sc FIT3D} y el obtenido mediante otros paquetes de uso frecuente, encontrando que los par\'ametros derivados son totalmente compatibles.
}

\addkeyword{Galaxies: ISM - Galaxies: ISM - Techniques: Spectroscopy}
\begin{document}
%\[% Typeset article header
\maketitle

\section{Introduction}
\label{intro}

The optical spectrum of a galaxy, or a part thereof, comprises information 
about the different components that emit or absorb light. Therefore, it can be
considered as the net sum of different emitting sources, mostly stars
and ionized gas, red- or blue-shifted and broadened by their
particular kinematics, and attenuated by dust. All these components
come together, co-added, to conform the observed spectrum, so they
have to be decoupled in order to study their individual properties.

Fortunately, some of them present clear observational
differences. While the stellar populations dominate the continuum
emission, the ionized gas shines in a set of clearly defined emission
lines at fixed rest-frame wavelengths defined by atomic
physics. Several different tools have been developed to model the
underlying stellar population, effectively decoupling it from the emission lines
\citep[e.g.,][]{cappe04, cid-fernandes05, ocvrik2006, sarzi2006, 
sanchez07a, koleva2009, macarthur2009, walcher11, sanchez11,wilk15}. 
Most of these tools are based on the same principles. It is assumed that
the stellar emission is the result of a single or a combination of
different single stellar populations (SSP), or resulting from a
particular star formation history (SFH).  It is well known that the
spectral energy distribution (SED) of simple stellar populations
(chemically homogeneous and coeval stellar systems) depends on a set
of first principles (e.g. initial mass function, star formation rate,
stellar isochrones, metallicity, etc.), from which it is possible to
generate the spectra of synthetic stellar populations. This technique,
known as evolutionary synthesis modeling
\citep[e.g.][]{Tinsley:1980p3431}, has been widely used to unveil the
stellar population content of galaxies by reconciling the observed
spectral energy distributions with those predicted by the theoretical
framework.

Unfortunately, the variation of different physical
quantities governing the evolution of stellar populations produce
similar effects in the integrated light of those systems, leading to a
situation in which the observational data is affected by hard to solve
degeneracies, such as the ones involving age, metallicity, and extinction 
\citep[e.g.][]{Oconnell:1976p3432,Aaronson:1978p3433,Worthey:1994p3434,gildepaz02}.
Even so, the use of spectrophotometric calibrated spectra and the
sampling of a wide wavelength range helps  breaking the degeneracy, and
allows the derivation of reliable physical parameters by fitting the full
spectral distribution with single stellar populations
\citep{Cardiel:2003p3435}.

However, the simple assumption that a single stellar population
describes well the SED of a galaxy is not valid even for early-type
galaxies, less so for late-type ones. Most galaxies present complex
star formation histories, with different episodes of activity, of
variable intensity, time scales, and complex dust
distributions. Therefore, a single stellar population does not
reproduce well their stellar spectra. A different technique, known as
full spectrum modeling, involving the linear combination of multiple
stellar populations and the non-linear effects of dust attenuation,
has been developed to reconstruct their stellar populations
\citep[e.g.][]{pant2003,cid-fernandes05,toje2007,ocvrik2006,sarzi06,koleva2009,macarthur2009}.

In general, these reconstructions require a wide wavelength range to
probe simultaneously the hot, young stars and the cool, old stars. 
They also require the best spectrophotometric
calibration to disentangle the effects of age, metallicity, and dust
attenuation. Although the different implementations of this technique have
some differences, they are very similar in their basis. The 
information extracted from the multi stellar population modeling differs among
 implementations. In some cases, the luminosity (or mass)
weighted ages and metallicities are derived, based on the linear
combination of different models \citep[e.g.][]{sarzi2006}. In other
implementations  the fraction of light (or mass) of
different stellar populations  is derived
\citep[e.g.][]{Stoklasova:2009p3436,macarthur2009}. Finally, in other
tools the information in the shape of the stellar continuum is
considered unreliable and removed prior to any further analysis
\citep[e.g.][]{ocvrik2006}.

In addition to the actual composition of the stellar population, 
kinematics effects need to be taken into account. The stellar
continuum should be redshifted by a certain velocity,
broadened and smoothed to account for velocity dispersion, and
attenuated due to a certain dust content. Some of the 
algorithms mentioned above perform this kinematic analysis prior to the
decomposition of the underlying stellar population
\citep[e.g.][]{cid-fernandes05}, while in other algorithms it is a
fundamental part of the computation \citep[e.g.][]{sarzi2006}.

Once the modelled stellar spectrum is determined, it is subtracted from the 
observed spectrum to provide a pure emission line spectrum that includes
the information from the ionized gas. In some algorithms the stellar 
and ionized gas spectra are analyzed together, although in many
cases a two-step procedure is used in the analysis \citep[e.g.][]{patri14a}. 
Then, the main properties of the emission lines, including the intensity and
kinematics, need to be derived. In general, given their larger signal to noise, 
it is assumed that the information
derived from emission lines is more accurate and stable than that
derived for stellar populations \citep[see][for a review of the state 
of the art]{walcher11}. However, a proper subtraction of the
stellar contribution is required prior to measure emission line ratios, such as
the [OIII]/H$\beta$ ratio, that allow to interpret the data in
terms of physical processes.

 In \citet{sanchez06b} we presented a new tool to perform all these
analyzes, mostly focused on the spectra obtained using Integral Field
Spectroscopy (IFS), in the optical spectral range. This tool, named {\sc FIT3D},
was first developed with the aim of analyzing the properties of the ionized
gas through its emission lines, and has been used in several science publications
and PhD theses. In the last few years {\sc FIT3D} has
increased its capabilities to derive a more reliable characterization
of the underlying stellar population, and a good estimation of the errors
of the parameters derived. 

This article is the first of a series focused on the description of
{\sc Pipe3D}, a spectroscopic analysis pipeline developed to
characterize the properties of the stellar populations and ionized gas
in the spatially resolved data from optical IFU surveys, in particular
CALIFA \citep{sanchez12b}, Mange \citep{manga}, and SAMI
\citep{sami}. {\sc Pipe3D} uses {\sc FIT3D} as its basic fitting
package. In this article we present the new algorithms included in the
distributed version of {\sc FIT3D}. In Section \ref{fitting} we
describe the new fitting philosophy and the details of the algorithms
implemented. In Section \ref{ac}, we demonstrate, based on extensive
simulations, the accuracy { and precision} of the parameters
recovered for the stellar populations (Sec. \ref{ac_ssp}), using a
simulated single-burst stellar population (Sec. \ref{rBurst}), or a
fully simulated star formation and chemical enrichment history
(Sec. \ref{rSFH}). The reliability of the errors estimated for the
parameters derived is described in Section \ref{rSSP_ER}.  Tests on
the accuracy { and precision} of the parameters recovered for the
emission lines and of the estimated errors are discussed in Sections
\ref{rEL} and \ref{rEL_ER}, respectively.  In Section \ref{simReal} we
explore the precision in the parameters derived, for both the stellar
populations and the ionized gas, with more realistic simulations based
on actual IFU data extracted from the CALIFA survey. A comparison with
more broadly used fitting algorithms is presented in Section
\ref{comp}. Finally, Section \ref{summ} summarizes the main
conclusions of this study.

\begin{figure*}[!t]
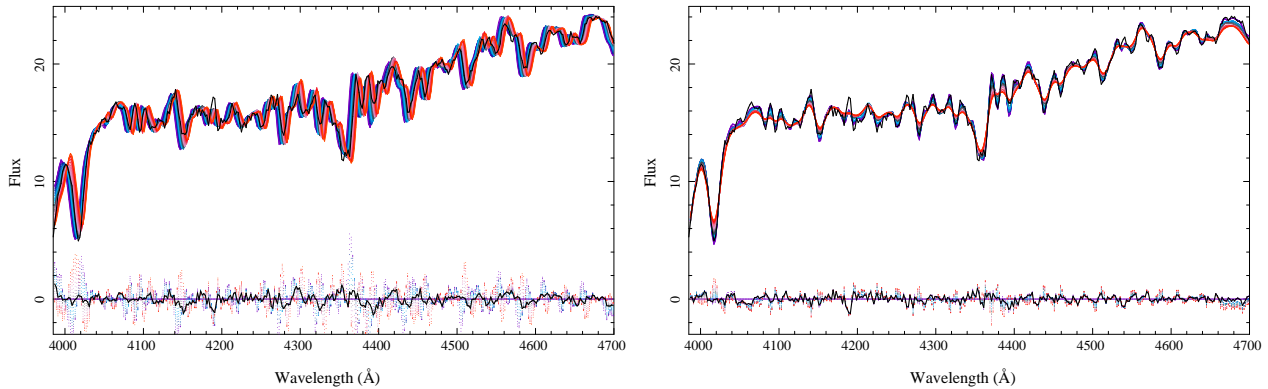

  \includegraphics[angle=270,width=0.5\linewidth]{figs/FIT3D_vel.ps}\includegraphics[angle=270,width=0.5\linewidth]{figs/FIT3D_sigma.ps}
  \caption{{\it Left Panel:} Detail of the central spectrum (5$\arcsec$ aperture, black solid-line) of NGC 2916 extracted from the V500 data cube of the CALIFA survey, together with the best fitted SSP template for the different scanned velocities (color-coded solid lines). The blueish solid lines correspond to SSP models blue-shifted with respect to the velocity, and the reddish solid lines correspond to red-shifted ones. {\it Right Panel:} Same spectrum shown in the left panel (black solid-line),  together with the best fitted SSP template for the different scanned velocity dispersions (color-coded solid lines). The blueish solid lines correspond to SSP models with smaller velocity dispersion, while the reddish solid lines correspond to those with larger velocity dispersion. In both panels the residuals from the fits are plotted using the same color scheme, together with the residual of the best fitting model (grey solid-line). { The two panels show two consecutive steps in the fitting procedure.}}  \label{fig:vs}
\end{figure*}

\section{New Fitting philosophy of FIT3D}
\label{fitting}

 The main goal of any analysis pipeline dealing with IFS galaxy data
 is to disentangle the main components comprising the spatially
 resolved spectra. As in the case of long-slit or single-fiber
 spectroscopic data, the two main components are: (i) the light coming
 from the stellar population, that dominates the continuum and
 absorption spectrum, and (ii) the emission of the ionized gas, that is
 observable as a set of emission lines at certain characteristic
 wavelengths. The two components are affected by dust attenuation
 ($A_V$), produced by dust grains  distributed across the
 galaxy \citep[e.g.][]{tuffs2004}. The disentanglement of these two
 main components is mandatory to properly study the properties of
 local and distant galaxies.

 The rest-frame spectra of the two components are observed redshifted due
the Doppler effect induced by the expansion of the universe and the residual
velocity of the galaxy, and further blue- or red-shifted by the particular kinematics
of the target cell, and component, within the galaxy.
In addition, the spectra are broadened by a characteristic velocity dispersion,
accounting for the unresolved kinematics that comprises the
perturbations with respect to the average velocity pattern, and the
non-homogeneous motions (such as non-circular motions). In rotational
supported systems, like disks of spiral galaxies, the velocity
dispersion is much lower than the asymptotic rotation velocity. On the
other hand, in pressure supported systems, like bulge dominated
early-type galaxies, the broadening from random orbits is much larger
than pure rotation velocity. Obviously, this picture is an oversimplification,
since many other components may affect the kinematics pattern, like
bars, outflows, interactions, an so on \citep[e.g.][]{jkbb14}.
Irrespectively of the particular kinematic pattern, its effect has to
be taken into account by any analysis pipeline.

Former versions of {\sc FIT3D} address this problem by (i) modeling the
continuum with a linear combination of synthetic stellar populations
(SSP), and (ii) modeling the emission lines with a set of single
Gaussian functions. Details of the  procedures adopted to fit both
components are given in the sections below. In the case of the
stellar populations, their weights or eigen-values were derived without
boundary constraints
\citep{sanchez06b,sanchez11}. However, if any of the  weights derived
was found to be negative, the corresponding SSP was removed from the
template list, and the fitting procedure was repeated until only positive
weights were derived. Under this procedure, only a  few number
of SSPs remain in the final iteration, limiting the extracted
information to the average properties of the underlying stellar
population, like the luminosity or mass weighted ages and
metallicities. This procedure was also very sensitive to the effects
of noise, and did not allow to estimate the errors on any of the
quantities derived. Despite these limitations, this approach provides
a reliable subtraction of the underlying stellar population.

In the current version of {\sc FIT3D}\footnote{http://www.astroscu.unam.mx/$\sim$sfsanchez/FIT3D}, 
we adopt a Monte-Carlo approach
(MC), in which the previous procedure is iterated for a randomized
version of the input spectrum taking into account the error vector.
The individual weight vector is stored in
each of the iterations. The final model of the stellar component is
built using the mean weights derived along the MC sequence. In
addition, this method provides realistic uncertainties of both the weights
and the final model, based on the standard deviation with respect to
the mean values. The details of how the MC is performed will be
explained in the following sections.

Therefore, the analysis of the underlying stellar population, without
considering the emission lines, comprises a non-linear and a linear
problem. The non-linear problem includes the parameters that define
the kinematic properties of the spectra and the dust attenuation, and
the linear one involves the decomposition of the underlying stellar
population in its components, either synthesized stellar populations
(SSPs), or a library of stellar templates. The non-linear
problem is addressed first and independently than the linear one.

{ The algorithms included in {\sc FIT3D} were originally developed
  in {\sc perl}, making an extensive use of the Perl Data Language
  ({\sc PDL }\footnote{http://pdl.perl.org/}). We have developed a
  {\sc python} version for which we have adopted most of the required
  algorithms included in {\sc PDL}. The particularities of the {\sc
    python} version will be described elsewhere (Ibarra et al., in
  prep.). }

\subsection{Non-Linear parameters of the stellar populations}
\label{sec:non_linear}

Unlike other tools \citep[e.g., {\sc pPXF}][]{cappellari04}, {\sc
  FIT3D} was not developed to provide a detailed kinematic analysis of
the line-of-sight velocity distribution. However, it estimates the
main properties of the kinematic structure, i.e., the stellar 
velocity ($v_{sys}$) and velocity dispersion ($\sigma_{v}$). These
parameters are derived under the assumption that all stellar
populations move at the same velocity, with a velocity dispersion
following a Gaussian function. This is clearly an approximation, since
it is well known that the velocity distribution of stellar populations
deviate strongly from this simple approach
\citep[e.g.][]{rix92}. However, most cases require a higher spectral
resolution than the one presented here ($R<$2000) to make a clear
distinction between this Gaussian approximation and more complex
velocity distributions. The stellar kinematics is derived prior to the
linear combination required to analyze the stellar population, based
on an additional pseudo-random exploration within a pre-defined range
of values.  The pseudo-random exploration varies the parameter in question
within a pre-defined range of values, but not following a fixed step. 
Instead it uses a random step, generated within a $\pm$50\% of the original
step (by default, 1/30th of the parameter range).
This loop is repeated several times (three by default). 
We consider that this non-regular exploration of the parameter space 
provides a better sampling within the given range of values.

For the velocity dispersion we adopted two different implementations: (i)
measuring it in units of wavelength, since
for most of the  spectra analyzed the dispersion is dominated by the instrumental
one, and (ii) a different version that considers
both a fixed instrumental dispersion along the spectral range in units
of wavelength, and a physical velocity dispersion (in km/s). That
second version requires a better knowledge of the instrumental
dispersion and it is somewhat slower to run since it performs
two different convolutions over the stellar templates. 

In the first case, the stellar templates are convolved with a Gaussian
function in wavelength space, and then shifted to the
corresponding velocity. Although this is a crude approximation, for
spectra dominated by instrumental resolution 
\citep[e.g., PINGS and F-CALIFA data][]{rosales-ortega10,marmol-queralto11}, 
there is no significant difference.  In the second
case, the stellar templates are first convolved in linear wavelength space
with a fixed instrumental resolution (as in the first case), and then they are
convolved in log($\lambda$) space with a Gaussian that accounts for
both the velocity and the velocity dispersion.

 The fitting sequence is as follows: First, the velocity
dispersion is fixed to an input guess value, and the velocity is
changed incrementally from a minimum to a maximum value, using a
semi-random step around a certain fixed velocity step. For each
velocity a best linear combination of SSPs that
reproduces the continuum is derived, based on the reduced $\chi^2$. Then a range
of velocity dispersions is explored, following a similar procedure,
now fixing the velocity to the best result found in the previous step.
The procedure is repeated twice, with the second iteration focused on
the best parameters found in the first one, and reducing both the
range and the size of the step to a 10\% of the initial values.
Figure \ref{fig:vs} illustrates this process. It shows a detail of
the central spectrum (5$\arcsec$ aperture), extracted from the V500 setup
data cube of NGC 2916 from the CALIFA survey, together with the best
SSP templates derived for each of the scanned velocities and
velocity dispersions within the ranges considered.

\begin{figure}[!t]
  \includegraphics[angle=270,width=1.05\linewidth]{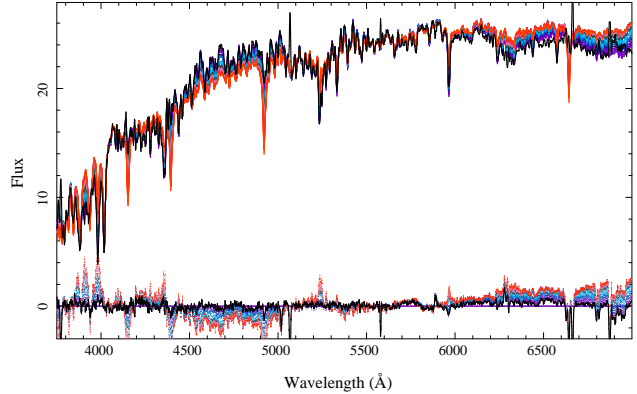}
  \caption{Central spectrum (5$\arcsec$ aperture, black solid line) of NGC 2916 extracted from the V500 data cube of the CALIFA survey, together with the best fitted SSP template for the different scanned dust attenuations ($A_V$) as color-coded solid lines ranging from $0$ to $1.6$ mag. The blueish solid lines correspond to SSP models with low dust attenuation, while the reddish solid lines correspond to those with larger attenuation. The residuals from the fits are plotted using the same color scheme, together with the residual of the best fitting model (grey solid line).} \label{fig:AV}
\end{figure}

Once the velocity and velocity dispersion are derived, then 
the range of dust attenuations that better reproduce the
data is explored. Again, a range of values is explored between a minimum and a
maximum dust attenuation, using a pseudo-random exploration with a
variable step between consecutive values. Following the same procedure
described before, the best linear combination of SSPs is derived 
for each particular value of the dust attenuation. Figure \ref{fig:AV}
illustrates this procedure. It shows the full wavelength range of the
central spectrum (5$\arcsec$ aperture), extracted from the V500 setup
data cube  of NGC 2916 from the CALIFA survey, together with the best
SSP templates derived for each of the scanned values of dust attenuation
within a range between 0 and 1.6 mag.

Obviously, as described, this procedure would be computationally very
intensive, and impractical in its current form. Thus we have
implemented some modifications to speed-up the process: (i) The MC
iterations on the linear combination of SSPs are not applied in the
steps required to derive the velocity, velocity dispersion, and dust
attenuation, thus, a single linear combination is derived for each
step. { The fitting is performed using a standard matrix inversion method to do a least squares minimization
weighted by the errors, as implemented in the {\sc PDL} routine {\sc linfit1d}\footnote{http://pdl.perl.org/PDLdocs/Fit/Linfit.html}}; (ii) a library with fewer number of single SSPs { was} adopted
in these initial steps too { (although it is not mandatory or restricted to do so)}; (iii) a restricted wavelength range could
be considered for the derivation of the kinematic components, taking
into account those regions with stronger stellar absorption features
(e.g., 3700-4700~\AA\ in the case of CALIFA and MaNGA data). This
approach is new, since in general it was considered that using just
the regions with stronger stellar absorptions would create a bias
towards the velocity of the younger stars. Indeed, H-K Ca lines can be
broadened due to the rotation of hot stars
\citep[e.g.][]{gerh2003,walcher05}. However, for the range of velocity
dispersion expected in galaxies this seems not to have a strong
effect, based on the tests that we have performed; (iv) a well defined
range for the velocity and velocity dispersion is derived by analyzing
the integrated spectrum and the central/peak spectrum for each
galaxy. In this initial exploration a wider range of parameters is
considered. Extensive tests have shown that these approaches do not
affect the results within the expected uncertainties, as we will
describe later.

In all the three pseudo-random explorations, for each parameter we consider
an input guess, a range of values, and an input step for which the parameter space is explored. 
The input guess is used as a fixed tracer of one of the parameters when 
 the other two are explored. Therefore, the accuracy { and precision} of the derivation is
limited to the  input parameters injected. In order to
perform the pseudo-random exploration, instead of using a constant step 
between the minimum and maximum, the step is variable within an
80\% range of the input step. This ensures that the exploration is not
uniform, and, due to the double iteration, it allows a better
estimation of the parameter. For each iteration the best
reduced $\chi^2$ derived from the linear analysis of the underlying
stellar population is stored. Then, the minimum $\chi^2$
together with the two adjacent values are used to derive the 
parameter value using a parabolic minimum derivation 
\citep{sanchez06a}, together with a rough estimation of the error in
the parameter as the range in which the reduced $\chi^2$
changes up to $\chi^2+1$.

%The emission lines are fitted using a somehow similar physolophy.
%Like in previous version of FIT3D the emission lines are fitted in
%each iteration of the processes using the gas-pure spectrum: i.e., the
%spectrum residual once subtracted the best model for the stellar
%population. As indicated before each emission line is modelled using a
%single or a set of gaussian functions.

%\section{Properties of the stellar population} 

\begin{figure*}[!t]
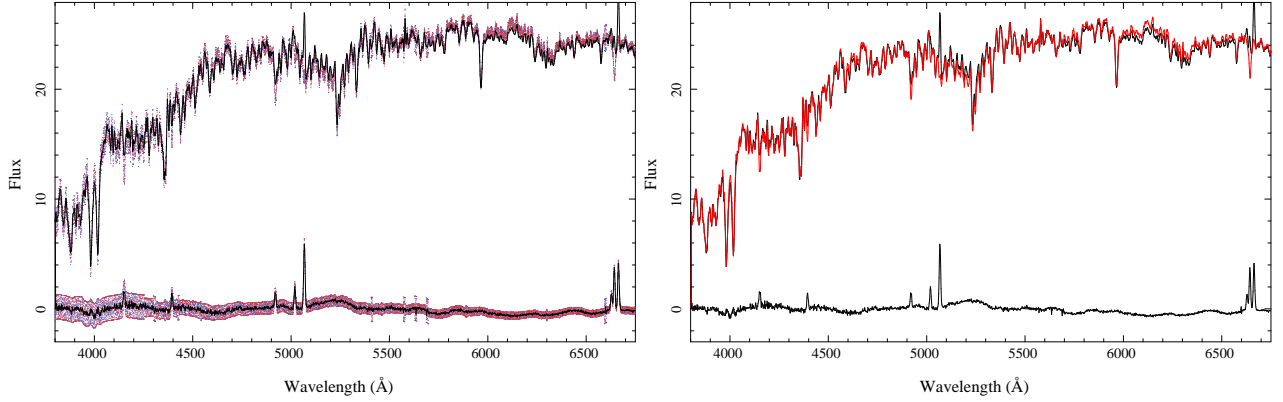

  \includegraphics[angle=270,width=0.5\linewidth]{figs/FIT3D_ORG.ps}\includegraphics[angle=270,width=0.5\linewidth]{figs/FIT3D_CAT.ps}
  \caption{{\it Left Panel:} Detail of the central spectrum (5$\arcsec$ aperture, black solid line) of NGC 2916 extracted from the V500 data cube of the CALIFA survey, together with the different MC-realizations of this spectrum created along the fitting procedure (color-coded points). The real noise has been increased by a factor 5 to visualize the different realizations more clearly. The residuals from the best model fitted in each MC-realization are included using the same color code as the model, together with the average of the different residuals (solid black line). {\it Right Panel:} Same spectrum shown in the left panel (black solid line) together with the best fitted SSP template derived using the MC linear regression technique (red solid line). The residual from the best fitting model is shown as a solid black line.}  \label{fig:MC_fit}
\end{figure*}

\subsection{Linear parameters of the stellar populations}
\label{sec:linear}

Once the non-linear parameters are derived, they are fixed in the
derivation of the properties of the underlying stellar population,
reducing the problem to the solution of a linear combination of
templates (SSPs).

The implementation of the multi stellar-population fitting technique
used in this article was already described in
\citet{sanchez06b,sanchez11}. The basic steps of the fitting algorithm,
spectrum by spectrum, are the following:

\begin{enumerate}

\item Read the input and noise spectra and determine the areas to be masked. Lets define $O_i$ as the observed flux at a certain wavelength $\lambda_i$, $\sigma_i$ as the noise level at the same wavelength, and $N$ the number of elements of the masked spectrum. 

\item Create a Monte-Carlo realization of the data, adopting a Gaussian random distribution for the noise. Lets define $G_i$ as the MC-realization
of the observed flux, defined as

$$ G_i^k = O_i + {R_i}\sigma_i$$

where ${R_i}$ is a random value following a { (-1,1)} Gaussian distribution, 
and $k$ is the running index of the current MC realization.

\item Read the set of SSP template spectra, shift them to the 
  velocity of the  spectrum, convolve them with the velocity dispersion, and re-sample to the wavelength
  solution. Lets define $F_{ji}$ as the flux of the $i^{th}$ wavelength of the
  $j^{th}$ template (once shifted, convolved, and re-sampled), where $M_0$ is the
  total number of templates used.

\item Apply the derived dust attenuation to the templates. The attenuation law of
  \citet{cardelli89} was adopted, with a ratio of total to
  selective attenuation of $R_V$=3.1.  Let us define
  $F_{ji}^{A_V}$ as the flux of the $i^{th}$ wavelength of the  $j^{th}$
  template, after applying the dust attenuation corresponding to a certain
  attenuation of $A_V$ magnitudes.

\item Perform a linear least-square fit of the input spectrum with the
  set of redshifted, convolved, and dust attenuated SSP templates.
  { The fitting is performed using a standard matrix inversion
    method, weighted by the errors, as described before, based on the
    {\sc PDL} routine {\sc linfit1d}. Then it is adopted} a modified
  $\chi^2$ as the merit function to be minimized, with the form:

$$ \chi^2 = \frac{1}{N-M} \sum_{i=1}^{N} D^2_i, $$

\noindent where

$$ D_i = w_i \Big( G_i^k- \sum_{j=1}^{M} a_j F_{ji}^{A_V} \Big), $$

In the above expression, $w_i$ is the weight of the $i^{th}$ pixel, defined as

$$ w_i = \frac{1}{\sigma_i^2}, $$

\noindent $a_j$ is the coefficient of the $j^{th}$ template in the
final modeled spectrum, and $M$ is the number of templates considered
in the fitting procedure. { The actual number of templates included
  in the SSP-library depends on the science goals and the quality of
  the explored data.  In our case for a typical $N\sim$2000, the
  number of templates if of the order of $M\sim$100. There is no
  general recommendation of how large should be $M$ to derive optimal
  results. It is therefore required to perform simulations to test it
  \citep[e.g.][]{cid-fernandes14}. However, our experience indicates
  that it should not be larger than 2-3 types the typical
  signal-to-noise of the spectra, as a practical rule.}

\item Determine for which templates within the library a negative
  coefficient is derived in the linear combination (i.e. $a_j<0$). These templates will
  be excluded from the next iteration of the fitting procedure, that will
  be resumed in step (5). At each iteration, $M$ is decreased by the amount of
  templates excluded. This loop ends once all the coefficients are positive.
  { It is important to note here that excluding those templates within the library
  that produce negative values for the coefficients do not produce in general 
  an increase of the reduced $\chi^2$ of the fitting. Therefore, to remove them 
  do not decreases the goodness of the fitting in a systematic way. This procedure
  is adopted both in the linear and non-linear exploration of the parameters.}

\item The modeled spectrum flux at $\lambda_i$ for the $k$th MC realization is given by

$$ S_i^{k} = \sum_{j=1}^{M} a_j^{k} F_{ji}^{A_V}, $$

\noindent where now $M$ considers only the templates with positive coefficients $a_j^k$.

\item The procedure is iterated for a fixed number of MC realizations ($K$) from
step (ii). The final modeled spectrum, with its corresponding uncertainties at each $\lambda_i$, is given by
the mean and standard deviation of the individual models for each realization of the MC iteration, 
at the given wavelength:

$$ S_i = {\rm mean} (S_i^{1..K})  $$
$$ \sigma_{S_i} = {\rm stddev} (S_i^{1..K})  $$

{ The final reduced $\chi^2$ is then recomputed using this final average modeled spectrum. It is found
that in general this $\chi^2$ is not significantly different that the individual ones for
each of the MC realizations (the same is valid for the standard deviation, in a consistent way).
This confirms the assumption that each individual fitting reproduce the analyzed spectrum equally well,
and the final distribution is a more realistic description of the properties of the stellar population.}

The final coefficients of the SSP decomposition and the corresponding uncertainties are derived using
a similar procedure:

$$ a_j = {\rm mean} (a_j^{1..K})  $$
$$ \sigma_{a_j} = {\rm stddev} (a_j^{1..K})  $$

where $\sigma_{par}$ is the uncertainty associated with the corresponding parameter.

\end{enumerate}

For practical reasons, the algorithm provides the fraction of the
total flux contributed by each particular SSP within the library at a
certain wavelength, and therefore a normalization wavelength is {
  recommended}, and the flux at that wavelength, to use the
coefficients for further derivations { in a more easy way}.  Let's
define $c_j$ as these relative flux fractions. { Along this article
  we have adopted the normalization at 5500\ \AA, however, this is not
  mandatory.}.

In the derivation of the non-linear parameters a reduced version of this algorithm was adopted,  
excluding steps (2), (6), and (8). Thus, only a single linear least-square fit  
is applied, without {\it cleaning} the negative coefficients (that are normally absent when the  template adopted
is limited to a few stellar populations), and not iterating over a sequence of MC realizations.

This algorithm shares a basis with other procedures described in the literature
\citep[e.g.,][]{cappellari04,cid-fernandes05,ocvrik2006,sarzi2006,walch06,sanchez07a,koleva2009,macarthur2009}. 
If required, low frequencies in the spectrum can be
fitted by adding  a polynomial function to the fitted templates, or by
multiplying them by that function. There are different origins for these
low frequencies in the spectrum, mostly related with defects in the spectrophotometric
calibration of the data or the stellar libraries adopted for the SSP templates \citep[e.g.][]{cid-fernandes13}
They may also be related to the way that the synthetic populations are generated, like
differences/errors in the adopted IMF or the stellar evolution models, being different
for different stellar templates \citep[e.g.][]{cid-fernandes14}. { In the
particular implementation shown along this article we have performed this correction 
only to subtract the stellar component to analyze the emission lines. However, for the
full analysis of the stellar populations is not implemented. This is evident in the residuals shown
in Fig. \ref{fig:MC_fit}.}

By construction, the fitting algorithm requires that certain regions
in the spectrum are masked: 
(i) strong and variable night sky emission line residuals, 
(ii) regions affected by instrumental signatures, like defects in the CCD (e.g. dead columns) 
whose effect was not completely removed during the data reduction process, 
(iii) regions affected by telluric absorptions, not
completely corrected during the flux calibration process, 
and (iv) regions containing strong emission lines from the ionized gas.

In the case of FIT3D  a final iteration is implemented that makes
this later masking not critical. It is possible to combine the
analysis of the stellar population and emission lines fitting in a
single iterative process in which, for each iteration, the best model
of the emission lines is removed prior to the analysis of the 
stellar population. We will describe later this possibility.

\subsection{Average properties of the stellar populations}
\label{sec:average}

Once the best model for the underlying stellar population is  derived based
on a linear combination of SSPs, it is possible to derive the average
quantities that characterize this population. 
The most common parameters derived  are:

\begin{itemize}

\item The luminosity-weighted log age ($Age_L$) and log metallicity ($Z_L$) of the underlying
  stellar population,

$$ {\rm log}Age_L = \sum_{j=1}^{M} c_j {\rm log}Age_j, $$

\noindent and

$$ {\rm log}Z_L = \sum_{j=1}^{M} c_j {\rm log}Z_j, $$

\noindent where $Age_j$ and $Z_j$ are the corresponding age and
metallicity of the $j^{th}$ SSP template, and $c_j$ is the weight in
light of each SSP. { In this context we define $Z$ as the mass fraction of metals, i.e., 
elements heavier than Hydrogen and Helium.}.

Hereafter, for the sake of simplicity, we will refer to the previous values as
the 'luminosity-weighted age' (instead of the more cumbersome
luminosity-weighted logarithmic age). The reader should take into account that
a different definition would be:

$$ Age_L = \sum_{j=1}^{M} c_j Age_j, $$

\noindent and

$$ Z_L = \sum_{j=1}^{M} c_j Z_j, $$

\noindent being the case that the logarithm of the later ones do not correspond to the former ones.
Indeed the former ones are the geometrical average, weighted by the fraction of light
corresponding to each single stellar population, while the later ones correspond
to the arithmetic average. Since the time sampling of 
the stellar templates is logarithmic, we consider the former one a better representation
of the average ages and metallicities in galaxies, although
it is possible to derive any of them using the {\sc FIT3D} output.

\item The mass-weighted log age ($Age_M$) and log metallicity ($Z_M$) of the underlying
  stellar population,

$$ {\rm log}Age_M = \sum_{j=1}^{M} c_j {\rm ML}_j {\rm log}Age_j, $$

\noindent and

$$ {\rm log}Z = \sum_{j=1}^{M} c_j {\rm ML}_j {\rm log}Z_j, $$

\noindent where $Age_j$ and $Z_j$ are the same parameters described before,
and ${\rm ML}_j$ is the mass-to-light ratio of the $j^{th}$ SSP template (i.e., $\frac{M}{L}$).

\noindent Hereafter we will refer to them as the mass-weighted age and metallicity. 
Similar caveats should be applied to this definition as the one expressed
regarding the luminosity-weighted ones.

\item The average Mass-to-Light ratio,

$$ <{\rm ML}> = \sum_{j=1}^{M} c_j {\rm ML}_j  $$

that in general is not the same as the Mass-to-light ratio of the integrated
spectra, that would be given by:

$$ {\rm ML} =  \frac{\sum_{j=1}^M M_j}{\sum_{j=1}^M L_j}  $$

where

$$ M_j = c_j {\rm ML}_j L $$

and $L$ is the total luminosity of the spectrum.

\item We define the star formation history (SFH) as the evolution of the star-formation rate 
along the time. To derive it, we define the mass of stars of a given age:

$$ M_{age} = \sum_{j=1}^{M} {\rm M}_j  ~~{\rm if} ~~ age_j = age$$.

Now, we estimate the cumulative amount of mass up to a time $\tau_{age}$ where
$\tau_{age} = \tau_{Universe} - age$ and $\tau_{Universe}$ is the age of the Universe 
at redshift zero. For doing so, we integrate
the amount of stellar mass since the beginning of the cosmological time
to that time ($\tau_{age}$):

$$ M_{age,c} = \sum_{j=1}^{M}  {\rm M}_j  ~~{\rm if} ~~ age_j > age. $$

\noindent From these cumulative masses it is possible to derive the star-formation rate (SFR) at
any particular time,

$$ SFR_{\tau_{age}} = \frac{\Delta M_{age}}{\Delta age}, $$

\noindent the distribution of these SFR$_{\tau_{age}}$ conforms the SFH as defined before.

\noindent At any time the mass remaining in stars is derived by:

$$ M_{age,cor} = \sum_{j=1}^{M} f_{cor,j} {\rm M}_j  ~~{\rm if} ~~ age_j > age, $$

\noindent where $f_{cor,j}$ is the correction factor that takes into account the mass-loss
and mass lock into remnants, for each stellar population at the given age and metallicity \citep[e.g.][]{court13}.

\noindent For each spectrum, both the SFH and the SFR are not single parameters, but arrays of length the number
of ages included in the SSP template library.

 \end{itemize}

The luminosity-weighted ages and metallicities should be considered as
the first momentum of the distribution of weights, or, in other words,
the first momentum of the star formation and chemical enrichment
histories (in logarithm scale as defined here).  They differ
considerably from the {\it effective} ages and metallicities, since
they do not match in general with the corresponding values if the
population was composed by a single SSP \citep{serra07}.

The luminosity-weighted ages and metallicities highlight the contribution
of the young stellar populations (with a strong color effect), that
are much more luminous for a certain mass than their older
relatives. However, the mass-weighted ages and metallicities are less
sensitive to the young stellar populations, and therefore trace better
the bulk of the star formation history, rather than the more recent
star-forming events.

\begin{figure*}[!t]
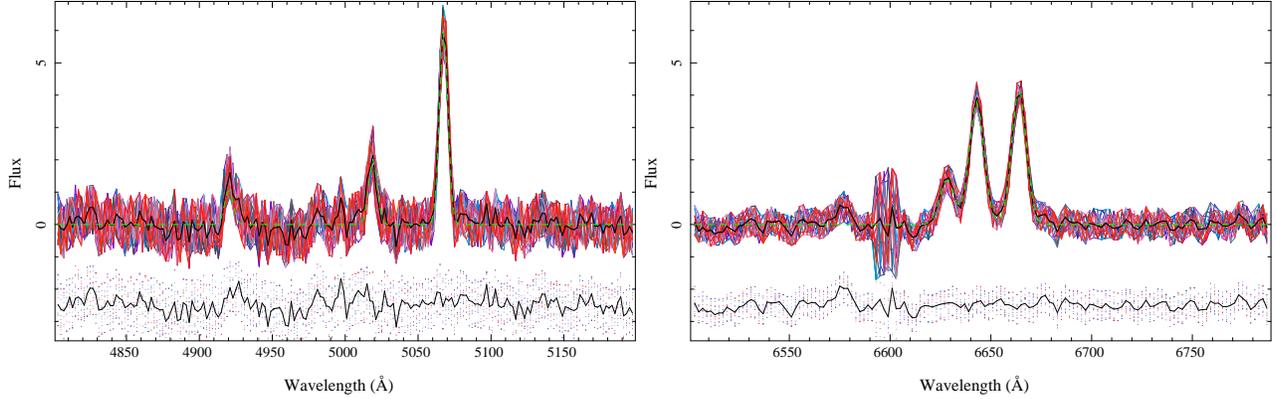

  \includegraphics[angle=270,width=0.5\linewidth]{figs/FIT3D_Hb.ps}\includegraphics[angle=270,width=0.5\linewidth]{figs/FIT3D_Ha.ps}
  \caption{{\it Left Panel:} Detail of central spectrum (5$\arcsec$ aperture, black solid line) of NGC 2916 extracted from the V500 datacube of the CALIFA survey around the wavelength range of H$\beta$ and  [\ion{O}{iii}]$\lambda$4959,5007, together with the different MC-realizations of this spectrum created along the fitting procedure (color-coded points). The best fitting model to the emission lines is shown as green dashed line. The residuals from the different realizations are shown following the same color scheme (offset by -2.5 flux units for the sake of clarity), together with the residual of the best fitting model. {\it Right Panel:} Same spectrum shown in the left panel, but now around the wavelength range of H$\alpha$ and  [\ion{N}{ii}]$\lambda$6548,6583 (black solid line),  together with the different MC-realizations of this spectrum created along the fitting procedure (color-coded points). The best fitting model to the emission lines is shown as a  green dashed line. In both panels the level of noise has been increased by a factor 5 to visualize the different realizations more clearly. }\label{fig:elines}
\end{figure*}

\subsection{Characterization of the emission lines}
\label{sec:elines}

One of the goals of this fitting procedure is to provide an
accurate characterization of the underlying stellar population, to subtract
this contribution from the original spectrum,  deriving an 
emission line only spectrum. This gas emission spectrum is derived by

$$ C_i = O_i - S_i. $$

To measure the intensity of each emission line detected, each emission
line in the {\it clean} spectrum ($C_i$) is fitted with a single
Gaussian function, plus a low order polynomial. Following the new
philosophy described below, the emission lines are fitted splitting
the non-linear and the linear components. As in the case of the
stellar population, the Gaussian function comprises two non-linear
parameters, the velocity and velocity dispersion, and a linear one,
the intensity. On the other hand, the polynomial function included to
fit the continuum only comprises a combination of linear
components. The procedure is very similar to the one outlined
before. First, we perform an pseudo-random exploration of the
non-linear parameters, within a range of values, starting with the
velocity, and then the velocity dispersion. An initial guess of the
velocity dispersion is used as a fix parameter in the exploration of
the velocity, and later the velocity derived is fixed in the
exploration of the velocity dispersion.  In each step of the
pseudo-random exploration a least-square linear regression is
performed to derive the linear parameters (i.e., the intensity of each
emission line and the coefficients of the polynomial function) { by
  using the same standard matrix inversion method, weighted by the
  errors, described before, based on the {\sc PDL} routine {\sc
    linfit1d}.}  The reduced $\chi^2$ is stored in each iteration {
  as a proxy of the the goodness of the fitting.}. The final best
combination of non-linear and linear parameters is derived on the
basis of the minimization of the reduced $\chi^2$.

We estimate the errors of the optimized parameters by
performing a MC simulation, where the original emission line only spectrum
is perturbed by a noise spectrum that includes both the original
estimated error and the uncertainties in the best fitted SSP model. Therefore,
the intensity fitted at each wavelength $\lambda_i$ in each MC realization is

$$ C_i^k = C_i + {R_i} \sigma_{{\rm modified},i}$$

\noindent where $i$ is the spectral pixel, ${R_i}$ is a random value following 
a { (-1,1)} Gaussian distribution, $k$ is the running index of the
current MC realization, and

$$ \sigma_{{\rm modified},i} = \sqrt{\sigma_i^2+ \sigma_{S_i}^2 }$$

\noindent where $\sigma_i$ is the original error (the one used in the
analysis of the SSP), and $\sigma_{S_i}$ is the standard deviation 
around the best SSP model, $S_i$ (i.e., the uncertainty in this
model){ , as defined in point 8 of Sec. \ref{sec:linear}}. Using this approach, we propagate the uncertainties of the
subtraction of the underlying stellar population to the parameters
derived for the emission lines.

 Instead of fitting all the wavelength range at once, for each spectrum  
we extract shorter wavelength ranges that sample one or a few
 emission lines. This allows us to fit the
 residual continuum with the most simple polynomial function (in general, of 1st or 2nd order),  
and simplify the fitting procedure. When more than one emission { line } is
 fitted simultaneously, their velocities and dispersions
may be coupled. Thus, they are forced to be equal in order to decrease the number of free
 parameters and increase the accuracy of the de-blending process (if
 required). This is particularly useful when the velocity dispersion
 is dominated by the spectral resolution, and for the recovery of the
 intensity of weak emission lines, with poorly defined kinematic properties. 
 Figure \ref{fig:elines} illustrates this procedure. It shows
 two spectral ranges centered in H$\beta$ and H$\alpha$, of the central
 spectrum of NGC 2916 shown in Fig. \ref{fig:vs}-\ref{fig:MC_fit},
 once subtracted the best fitted SSP model shown in
 Fig. \ref{fig:MC_fit}, together with the best fitted Gaussians
 to H$\beta$ and [\ion{O}{iii}]$\lambda$4959,5007, 
 and to H$\alpha$ and [\ion{N}{ii}]$\lambda$6548,6583, respectively. 
 The different MC realizations adopted during the
fitting procedure are shown to illustrate how the procedure evaluates the
parameters and errors at the same time.

A modeled emission line spectrum is created, based on the results
of the last fit, using only the combination of Gaussian functions. 
This spectrum is given by

$$ E_i = \sum_{k=1}^{L} B_{k} * {\rm Gauss}_{ik} (\lambda_k,\sigma_k), $$

\noindent where $L$ is the number of emission lines in the model, $B_k$ is
the integrated flux of the $k^{th}$ emission line, and ${\rm Gauss}_{ik}$ is the
corresponding normalized ($\lambda_k$,$\sigma_k$) Gaussian evaluated at the $i^{th}$ pixel.

Finally,  this modeled emission line spectrum is subtracted  
from the original spectrum ($G_i$), deriving an emission line free spectrum given by

$$ GF_i = G_i - E_i. $$

\noindent This spectrum may be used to model again the stellar
population, by applying the procedure described before, but without
masking the emission line spectral regions, as outlined in
Section \ref{sec:linear}. Actually, our experience is that adopting
a simple SSP template library for the initial iteration, and a more
detailed one for the second iteration produces very reliable results
for both the emission lines and the stellar populations. In general, a
third iteration does not produce any significant change in the properties derived, 
although the algorithm  implemented allows the user to
select the desired number of iterations. It is important to remark
that the final  reduced $\chi^2$ adopted includes both models and the
propagation of the uncertainties.

\begin{figure*}[!t]
  \includegraphics[width=\linewidth]{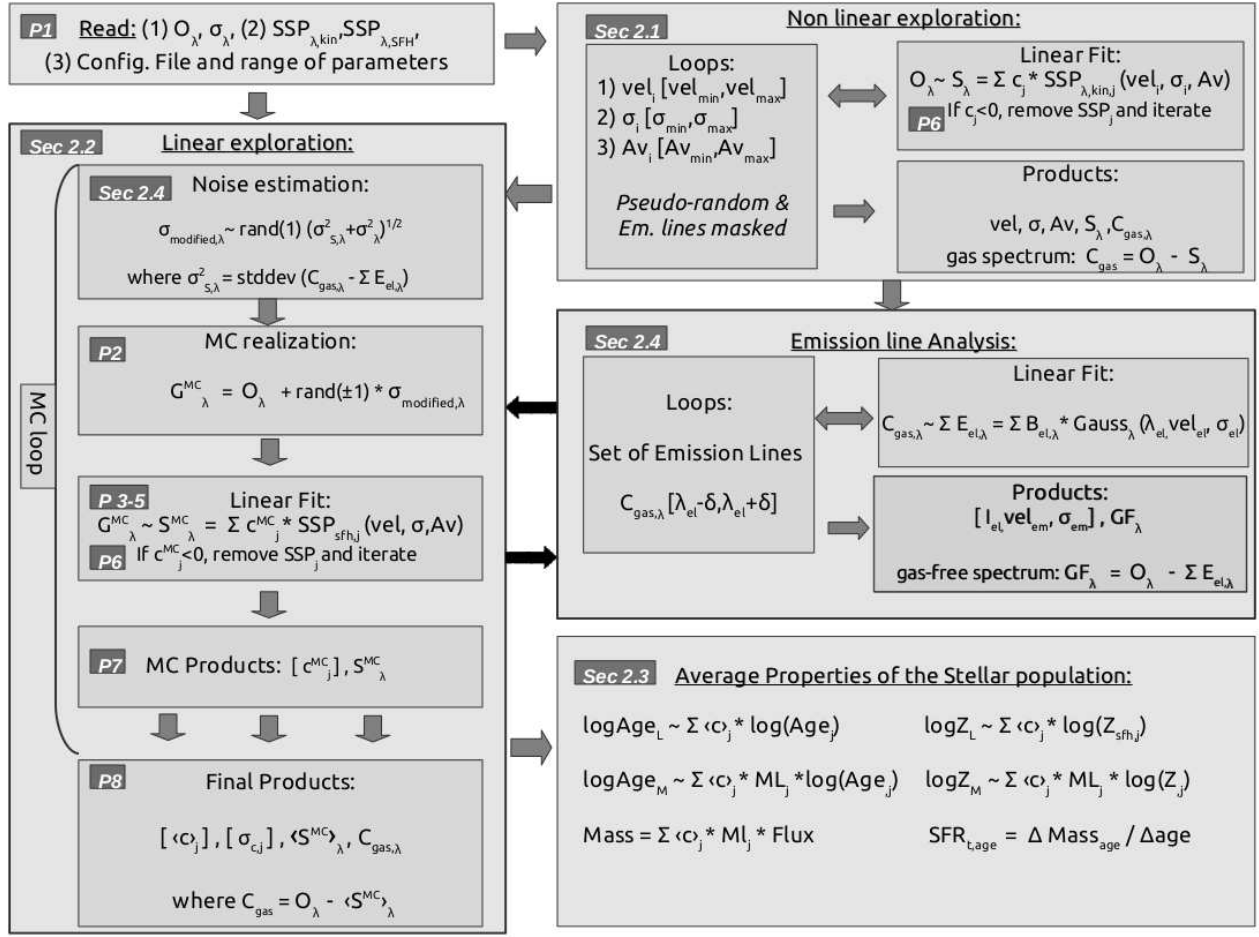}
  \caption{Chart-flow summarizing the fitting sequence of {\sc FIT3D} as described in the text. The flow indicate the main blocks of the fitting process, indicating which operations are performed and the main data-products derived. The MC-loop is indicated with a bracket, and the final iteration over the emission-line fitting is indicating with black arrows. In each box a label indicate either the section in the text where the procedure is described or the point in Sec. 2.2 in which is explained. }\label{fig:flow}
\end{figure*}

The full data-flow described in the previous sections is summarized in
Figure \ref{fig:flow}. In there, different steps of the fitting
sequence are described, illustrating the products derived in each of
them and the interdependence between those steps. The algorithm
implemented in {\sc FIT3D} allows the user to fit the stellar population
without fitting any emission line, and at the same time it includes
routines to fit the emission lines irrespectively of the underlying
stellar population, in case that this is required.

\section{Accuracy of the parameters derived}
\label{ac}

The procedure outlined above to fit the stellar population and the
emission lines implements a somewhat new concept.  It does not rely on
minimization algorithms used frequently to fit non-linear functions,
like the LM algorithm implemented in other tools (e.g., Gandalf, the
previous version of FIT3D), but on a sequential combination of
pseudo-random exploration of the range of non-linear parameters, and a
least-square inversion of the linear ones. The MC realizations and
pseudo-random explorations share some ideas with other tools that
explore the properties of the stellar populations, like {\sc
  Starlight} \citet{cid-fernandes05}, however, their minimization
scheme is based on a sequence of Markov chains, and not on a sequence
of linear regressions.

Therefore, we need to demonstrate that the new procedure derives
reliable results. To do so, we perform a set of simulations in
order to determine that (i) the adopted procedure creates accurate
models of the data, (ii) the main parameters of the stellar
populations and emission lines correspond at least to the expected ones
with respect to some well controlled inputs, and (iii) the procedure
provides a reliable estimation of the uncertainties of the parameters derived.

\subsection{SSP template library}
\label{ssp_lib}

As already noted by different authors
\citep[e.g.][]{macarthur+04, cid-fernandes14}, this kind of analysis
is always limited by the  template library adopted, that comprises a
discrete sampling of the SSP ages and metallicities. It is
desirable that the stellar library be as complete as possible, and
non-redundant. However, this would require an exact match between the
models and the data, which is not possible to achieve in general, 
in particular if the stellar population comprises more than one SSP.

The results should be as independent as possible of the currently
adopted SSP template library. Therefore, we have tested the fitting
routines using different ones, following \citet{cid-fernandes14}:

\begin{itemize}
 
\item {\tt gsd156} template library: described in detail by \citet{cid-fernandes13},
it comprises 156 templates that cover 39 stellar ages (1 Myr to 13 Gyr), and 4
metallicities ($Z/Z_{\odot}=$ 0.2, 0.4, 1, and 1.5). These templates were
extracted from a combination of the synthetic stellar spectra
from the GRANADA library \citep{martins05} and  the SSP library
provided by the MILES project \citep{miles, vazdekis10, falc11}. This library
has been extensively used within the CALIFA collaboration in different
studies \citep[e.g.][]{eperez13, cid-fernandes13, rosa14}. Therefore,
it is very interesting to know how accurate are our results using this particular
library. Its spectral resolution has been fixed to the spectral resolution
of the CALIFA V500 setup data (FWHM$\sim$6 \AA).

\item {\tt miles72} template library: It comprises 72
templates that cover 12 stellar ages (Myr to 12.6 Gyr), and 6
metallicities ($Z/Z_{\odot}=$ 0.015, 0.045, 0.2, 0.4, 1, and 1.5). These templates were
extracted from the SSP library
provided by the MILES project \citep{miles, vazdekis10, falc11}. They do not cover
as younger stellar populations as the previous ones, but they cover a  wider
range of metallicities. This template has a higher spectral resolution than the previous one,
with FWHM$\sim$2.5 \AA\ \citep{falc11}.

\item {\tt bc138} template library: It comprises 138
templates that cover 23 stellar ages (1 Myr to 13 Gyr), and 6
metallicities ($Z/Z_{\odot}=$ 0.005, 0.02, 0.2, 0.4, 1, and 2.5). 
The SSP models were created using the GISSEL code \citep{bc03}, assuming a
Chabrier IMF \citep{Chabrier:2003p3777}. They cover a similar range
of metallicities than the previous one, but with a wider range of ages. However,
they do not benefit from the spectrophotometric accuracy of the MILES
stellar templates \citep{miles}. Its spectral resolution is slightly higher than the {\tt gsd156} one,
with FWHM$\sim$5.5 \AA.

\item {\tt mar136} template library: It comprises 136 templates that
  cover 34 stellar ages (0.1 Myr to 15 Gyr), and 4 metallicities
  ($Z/Z_{\odot}=$ 0.05, 0.4, 1, and 2.23). The SSP models were created
  using the code by \citet{mar05}, assuming a Salpeter IMF
  \citep{Salpeter:1955p3438}. They cover a slightly smaller range of
  metallicities than the previous one, but with a similar range of
  ages. It is the template with the lowest spectral resolution, 
  FWHM$\sim$40 \AA. This template was included to test the ability to
  recover the stellar population parameters when the spectral
  resolution is not very accurate.

\end{itemize}

\begin{figure}[!t]
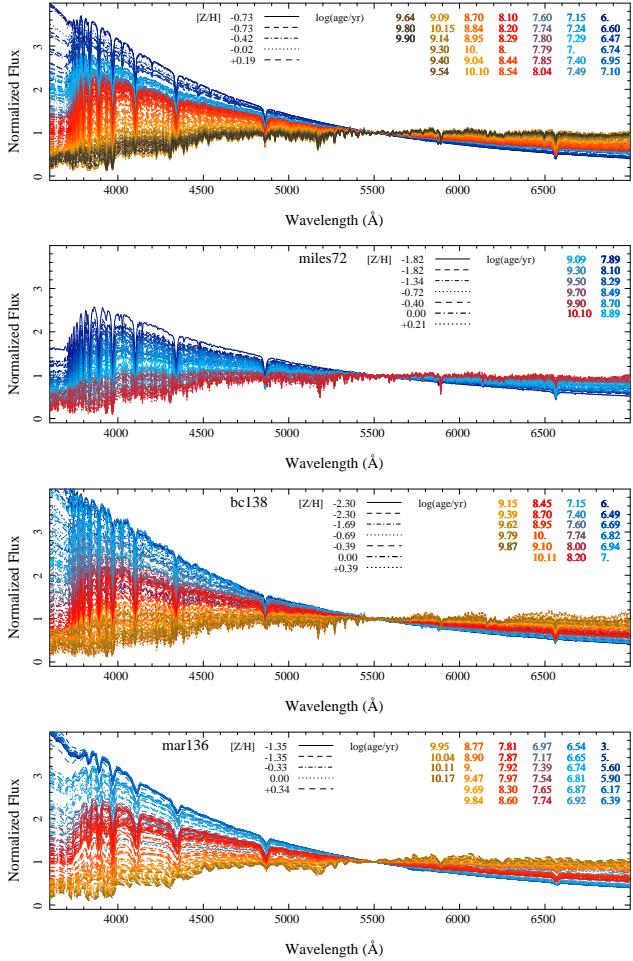

  \includegraphics[angle=270,width=1.05\linewidth]{figs/gsd61_156.ps}
  \includegraphics[angle=270,width=1.05\linewidth]{figs/miles_72.ps}
  \includegraphics[angle=270,width=1.05\linewidth]{figs/bc03_138.ps}
  \includegraphics[angle=270,width=1.05\linewidth]{figs/mar05_136.ps}
  \caption{Stellar population templates included in the  {\tt gsd156} (first panel), {\tt miles72} (second panel), {\tt bc138} (third panel) and {\tt mar136} (fourth panel) libraries. Each spectra within each library is represented with a different color, with the blueish colors representing the younger stellar populations, increasing the age towards more reddish/orange/grey colors.}
  \label{fig:ssp}
\end{figure}

The ranges of shapes covered by the different SSPs and how they
compare with each other is shown in Figure \ref{fig:ssp}. For each
SSP library we present the individual spectra within the wavelength
range corresponding to the optical range between 3700-7000\AA,
normalized at 5500\AA. We note that the sampling of all of these
libraries is different in the age distribution and in the metallicity
one. This is intrinsic to the templates, and imposes a bias in the
sampling of the properties, that is not intrinsic to {\sc
  FIT3D}. In addition, none of these libraries includes the nebular
continuum, to our knowledge, which may produce a significant effect
for very young stellar populations.

There are several caveats when applying model SSPs to the integrated
light of an entire galaxy, or a portion of it, which have been clearly
identified by previous authors
\citep[e.g.][]{macarthur+04, cid-fernandes14}. The most important one
is to assume that the parameter space covered by the empirical library
represents well that of the real data. However, in general, libraries
are based on stars in the solar neighborhood \citep[e.g.][]{miles} or
synthetic models \citep[e.g.][]{rosa2005, mar05}, and therefore it is
not granted that they represent well the stellar populations in other
galaxies (or even in other regions of our Galaxy). 
%There are other
%potential problems related to the particular selected templates.  For
%example, it is well known that the \citet{bc03} models have problems
%when dealing with the non-solar abundance ratios.
Most of these problems are not particularly important in the context of our tests,
since our primary goal is to determine the accuracy { and precision} of the 
parameters recovered for a set of simulated data; however, they do affect the
interpretation of the results.

\begin{figure*}[!t]
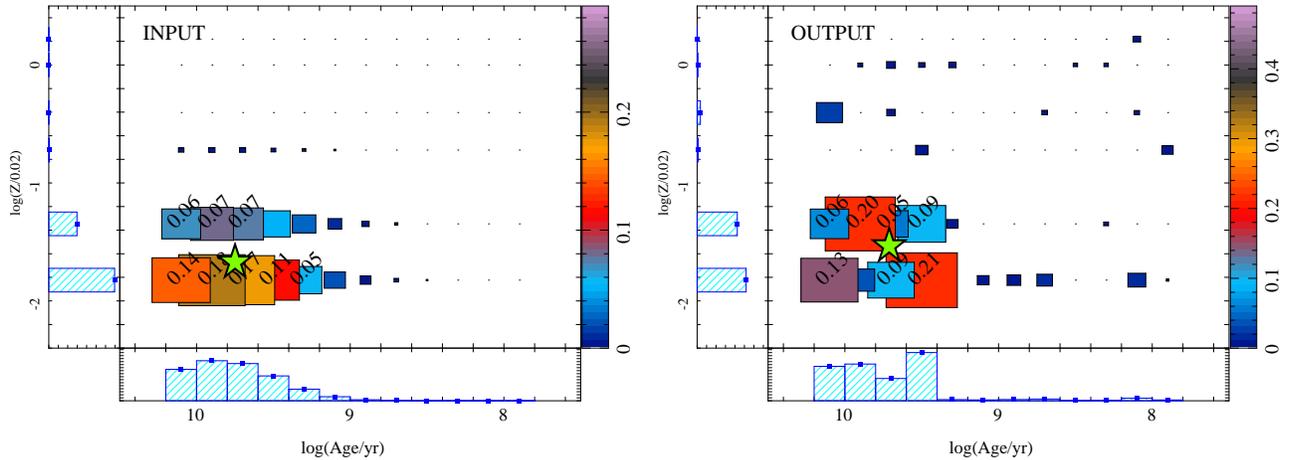

  \includegraphics[angle=270,width=0.51\linewidth]{figs/SFH_burst_IN.ps}\includegraphics[angle=270,width=0.51\linewidth]{figs/SFH_burst_OUT.ps}
  \caption{ {\it Left panel}: Distribution of input coefficients  along the age-metallicity grid for a { single} simulated spectrum corresponding to the first set of simulations, i.e., those resembling a single burst of star-formation, with a broad extension in time. This particular example corresponds to a simulation based on the {\tt miles72} template library. For each age-metallicity pair it is shown the weight as a solid square, with the size and color being proportional to the weight of the corresponding SSP$_{age,met}$. For those SSPs with stronger contribution (above a 5\%), the actual fraction is indicated on top of each square. The horizontal and vertical histograms represent the co-added coefficients for a particular age and metallicity respectively. { The green solid-star indicates the location of the luminosity weighted age and metallicity corresponding to the distribution of stellar populations: log(age/yr)=9.76, log(Z/0.02)=-1.68.} {\it Right panel:} Similar distribution corresponding to the  coefficients recovered for the given input simulation of a spectrum with a signal-to-noise ratio of $\sim$100, once the fitting procedure is  applied. { The green solid-star indicates the location of the luminosity weighted age and metallicity corresponding to the recovered distribution of stellar populations: log(age/yr)=9.78, log(Z/0.02)=-1.56.} }  \label{fig:Age_Met_burst}
\end{figure*}

In addition, it is important to note that the treatment of the dust
attenuation may affect the resulting parameters (i.e. the
luminosity-weighted age and metallicity of the stellar population). In
this particular implementation of the analysis we adopted the
\citet{cardelli89} attenuation law, which may not be the optimal
solution to study the dust attenuation in star forming galaxies
\citep[e.g.][]{calz01}.  Other authors, \citet{macarthur2009} adopted
a completely different attenuation law, based on the two-component
dust model of \citet{Charlot:2000p3439}, which is particularly
developed to model the dust attenuation in star forming galaxies. The
differences between the existing extinction laws in the optical
wavelength range (3600-10000 \AA) are too small to affect
significantly the
results\footnote{e.g., \url{http://www.mpe.mpg.de/~pschady/extcurve.html}}.
Indeed, no major differences are found when adopting the
\citet{cardelli89}, \citet{calz01}, or even a simple $\lambda^{-1.3}$
extinction law. {\sc FIT3D} adopted a fixed specific dust attenuation
of R$_V=$3.1 (i.e., Milky-Way like), and the only fitted parameter is
the dust attenuation at the $V$-band ($A_V$), in magnitudes.

%----------------------------------------------------------------
\begin{table*}[!th]\label{simtab}
\begin{center}
\caption[Results of the simulations: Single Burst]{ Results of the simulations: Single Burst.}
\begin{tabular}{rrrrcrr}\hline\hline
\multicolumn{6}{c}{Results using the {\tt gsd156} SSP template} \\\hline
\multicolumn{1}{c}{S/N} & \multicolumn{1}{c}{$\Delta$Age/Age} &
\multicolumn{1}{c}{$\Delta$Z/Z} & \multicolumn{1}{c}{$\Delta$A$_V$$^1$} &
\multicolumn{1}{c}{$\Delta$v$_{sys}$$^1$} & \multicolumn{1}{c}{$\Delta$$\sigma_v$$^1$} 
& \multicolumn{1}{c}{$r_{cor}$$^2$} 
\\\hline
%\tableline
64.4 &  0.17\,$\pm$0.11 & 0.07\, $\pm$0.10 & -0.06\,$\pm$0.17 &  0.6\,$\pm$ 40.9 & -34.4\,$\pm$39.3 & 0.38\\
32.9 &  0.16\,$\pm$0.21 & 0.05\, $\pm$0.19 & -0.08\,$\pm$0.18 &  2.5\,$\pm$ 49.5 & -41.2\,$\pm$45.4 & 0.38\\
 6.8 &  0.12\,$\pm$0.27 & 0.00\, $\pm$0.21 & -0.07\,$\pm$0.26 &-22.6\,$\pm$111.9 & -87.1\,$\pm$69.2 & 0.31\\
 3.4 &  0.09\,$\pm$0.31 & 0.00\, $\pm$0.24 &  0.05\,$\pm$0.32 &-38.1\,$\pm$208.40 &-104.4\,$\pm$74.1 & 0.24\\\hline
\multicolumn{6}{c}{Results using the {\tt miles72} SSP template} \\\hline
\multicolumn{1}{c}{S/N} & \multicolumn{1}{c}{$\Delta$Age/Age} &
\multicolumn{1}{c}{$\Delta$Z/Z} & \multicolumn{1}{c}{$\Delta$A$_V$} &
\multicolumn{1}{c}{$\Delta$v$_{sys}$$^1$} & \multicolumn{1}{c}{$\Delta$$\sigma_v$$^1$} 
& \multicolumn{1}{c}{$r_{cor}$$^2$}  \\\hline
%\tableline
70.6 &  0.07\,$\pm$0.15 & -0.05\, $\pm$0.31 & -0.02\,$\pm$0.17 &  -2.5\,$\pm$ 37.7 & -14.2\,$\pm$23.8 & 0.58\\
35.7 &  0.07\,$\pm$0.15 & -0.01\, $\pm$0.35 & -0.02\,$\pm$0.18 &  -6.2\,$\pm$ 43.3 & -19.5\,$\pm$31.1 & 0.57\\
 7.2 &  0.09\,$\pm$0.25 &  0.04\, $\pm$0.48 & -0.07\,$\pm$0.28 & -42.0\,$\pm$150.0 & -72.8\,$\pm$66.7 & 0.52 \\
 3.5 &  0.03\,$\pm$0.32 &  0.06\, $\pm$0.50 &  0.00\,$\pm$0.35 &-101.6\,$\pm$222.2 & -92.4\,$\pm$74.0 & 0.39\\\hline
\multicolumn{6}{c}{Results using the {\tt bc138} SSP template} \\\hline
\multicolumn{1}{c}{S/N} & \multicolumn{1}{c}{$\Delta$Age/Age} &
\multicolumn{1}{c}{$\Delta$Z/Z} & \multicolumn{1}{c}{$\Delta$A$_V$} &
\multicolumn{1}{c}{$\Delta$v$_{sys}$$^1$} & \multicolumn{1}{c}{$\Delta$$\sigma_v$$^1$}
& \multicolumn{1}{c}{$r_{cor}$$^2$}   \\\hline
%\tableline
65.4 &  0.09\,$\pm$0.50 & 0.17\, $\pm$0.64 & -0.10\,$\pm$0.35 &  6.7\,$\pm$ 37.1 & -55.7\,$\pm$45.3 & 0.44\\
34.3 &  0.13\,$\pm$0.48 & 0.16\, $\pm$0.66 & -0.12\,$\pm$0.35 &  0.1\,$\pm$ 79.0 & -63.8\,$\pm$48.8 & 0.44\\
 6.8 &  0.11\,$\pm$0.45 & 0.10\, $\pm$0.68 & -0.10\,$\pm$0.68 & -0.4\,$\pm$105.2 & -94.8\,$\pm$64.7 & 0.45\\
 3.4 &  0.04\,$\pm$0.44 & 0.15\, $\pm$0.75 & -0.03\,$\pm$0.34 &-15.1\,$\pm$113.3 &-108.7\,$\pm$65.2 & 0.47\\\hline
\multicolumn{6}{c}{Results using the {\tt mar136} SSP template} \\\hline
\multicolumn{1}{c}{S/N} & \multicolumn{1}{c}{$\Delta$Age/Age} &
\multicolumn{1}{c}{$\Delta$Z/Z} & \multicolumn{1}{c}{$\Delta$A$_V$} &
\multicolumn{1}{c}{$\Delta$v$_{sys}$$^1$} & \multicolumn{1}{c}{$\Delta$$\sigma_v$$^1$} 
& \multicolumn{1}{c}{$r_{cor}$$^2$}  \\\hline
%\tableline
66.0 &   0.04\,$\pm$0.30 & 0.25\, $\pm$0.41 & -0.15\,$\pm$0.28 &  16.9\,$\pm$ 59.0 & -77.7\,$\pm$72.8 & 0.53\\
35.6 &  -0.05\,$\pm$0.39 & 0.32\, $\pm$0.43 & -0.09\,$\pm$0.28 &  -2.0\,$\pm$ 75.2 & -91-3\,$\pm$70.3 & 0.54\\
 7.8 &   0.10\,$\pm$0.38 & 0.07\, $\pm$0.49 & -0.15\,$\pm$0.31 & -88.1\,$\pm$243.5 &-110.7\,$\pm$72.8 & 0.51\\
 3.5 &   0.16\,$\pm$0.43 &-0.01\, $\pm$0.53 & -0.22\,$\pm$0.36 &-134.1\,$\pm$260.4 &-117.8\,$\pm$76.0 & 0.57\\\hline
\multicolumn{6}{c}{Results using the {\tt gsd156} SSP template, fitted with the {\sc gsd32} template} \\\hline
\multicolumn{1}{c}{S/N} & \multicolumn{1}{c}{$\Delta$Age/Age} &
\multicolumn{1}{c}{$\Delta$Z/Z} & \multicolumn{1}{c}{$\Delta$A$_V$$^1$} &
\multicolumn{1}{c}{$\Delta$v$_{sys}$$^1$} & \multicolumn{1}{c}{$\Delta$$\sigma_v$$^1$} 
& \multicolumn{1}{c}{$r_{cor}$$^2$} \\\hline
%\tableline
63.6 &  0.14\,$\pm$0.20 &  0.08\, $\pm$0.17 & -0.08\,$\pm$0.19 &   3.9\,$\pm$ 25.8 & -35.0\,$\pm$38.0 &0.33\\
32.2 &  0.13\,$\pm$0.21 &  0.07\, $\pm$0.20 & -0.08\,$\pm$0.20 &   1.6\,$\pm$ 44.6 & -42.3\,$\pm$43.2 &0.29\\
 6.9 &  0.10\,$\pm$0.25 & -0.01\, $\pm$0.22 & -0.07\,$\pm$0.21 & -24.1\,$\pm$133.1 & -83.5\,$\pm$67.3 &0.19\\
 3.4 &  0.07\,$\pm$0.34 & -0.01\, $\pm$0.25 & -0.04\,$\pm$0.34 & -49.1\,$\pm$204.6 &-104.9\,$\pm$74.4 &0.17\\\hline
%32.19 &  0.45\,$\pm$0.68 & -0.07\, $\pm$0.28 & -0.17\,$\pm$0.49 & -11.6\,$\pm$ 72.5 & 38.4\,$\pm$99.5\\\hline
%734.60 & 0.14 &  0.02\,$\pm$0.14 & 0.0001\,$\pm$0.0014 & 0.00\,$\pm$0.05\\

%\tableline
% $^*$82.77   & 0.12 &  0.21\,$\pm$0.27 & 0.0010\,$\pm$0.0070&-0.01\,$\pm$0.10\\\hline
% $^{**}$81.97   & 0.13 &  0.12\,$\pm$0.47 & 0.0023\,$\pm$0.0075 &-0.01\,$\pm$0.23\\\hline
%\tableline
\end{tabular}
\end{center}

{ Each row shows the results of 1000 simulations, corresponding to a different
S/N level, for each particular stellar template. { Therefore, the errors in the derived offsets are a factor $\sqrt{1000}$ smaller than the values listed in the table.}}. $^{(1)}$ The dust attenuation $A_V$ is in magnitudes, while the units of the velocity and velocity dispersion are  km/s. 
$^{(2)}$ Correlation coefficient between $\Delta$Age/Age and $\Delta$Z/0.02.

%$^{**}$ Results when the input simulated spectra consists of a single SSP,
%instead of a combination of them.
\end{table*}
%----------------------------------------------------------------

\subsection{Accuracy of the properties of the stellar population}
\label{ac_ssp}
%\label{ssp_lib}

As mentioned before, the main parameters derived from the analysis of
the stellar component are (i) the coefficients of the SSP templates in
which the stellar continuum is decoupled, (ii) the luminosity- and
mass-weighted ages and metallicities, and dust attenuation, (iii) the
stellar mass-to-light and mass comprised in the spectrum, and (iv) the
star formation history. In order to assess the accuracy { and
  precision} of these parameters we have performed a set of
simulations. The simulations should be performed trying to match as
closely as possible some real data, especially the noise
pattern. Since the primary goal of the currently developed package is
the analysis of IFU data extracted from the CALIFA data, we will limit
our tests to the wavelength range and spectral resolution of the
CALIFA-V500 setup \citep{sanchez12a}: a wavelength range between
3700-7500 \AA, and an instrumental resolution of $\sigma_{inst}=$2.6
\AA. However, the simulations can be easily adapted to any other
spectroscopic configuration.

The noise pattern is composed, in general, of both white noise
(corresponding to the photon-noise of the source and the background,
and electronic noise from the detector), and non-white noise
(corresponding to defects/inaccuracies in the sky-subtraction,
uncorrected defects in the CCD, errors in the spectrophotometric calibration, etc). 
These noise patterns are
different spectrum-to-spectrum, and wavelength-to-wavelength, and are
clearly difficult to simulate on a simple analytical basis. However,
for a general test like the one performed here, we will limited our
simulations to the use of white noise. In forthcoming articles
we will explore the accuracy { and precision} of the parameters recovered  using a more {\it ad hoc}
set of simulations, following \citet{cid-fernandes14}.

We perform two different sets of simulations. In the first one, a SFH
dominated by a single burst of star formation is assumed, and
therefore the stellar populations are dominated by stars of a certain
age. The burst follows a Gaussian shape, with a width 0.5 dex in age
(i.e., the length of the burst is proportional to the age
selected). Under this assumption the burst lasts for a longer time at
earlier times, and it is much shorter at later times. For the
metallicity, we considered a similar probability
distribution. Therefore, the simulated stellar population is dominated
by stars of a certain metallicity with a Gaussian distribution of-width
of 0.5 dex at each metallicity.  This does not attempt to model any
real physical enrichment mechanism in particular.  Each simulated
stellar spectrum is created by randomly selecting (within the range of
the values considered in the SSP template) a particular age and
metallicity as the center of the 2D Gaussian distribution.

%Then it was selected a random
%value for the redshift, with in a range between $z\sim$0.005 and 0.03, the dispersion,
%within a range between 2 \AA and 8 \AA (corresponding to $\sim$100 and $\sim$300 km/s),
%and a dust attenuation between no attenuation and $A_v<$1.2 mag.

In general, this simulated SFH may not be realistic. In most 
 SFHs analyzed within the CALIFA survey most of the stars are formed
following an exponential law with more than 80\% of the
mass formed in the first few Gyr \citep{eperez13}. If we assumed this
more realistic SFH we would not be able to test the ability of the
fitting tool to decouple different stellar populations, and we would
just test the very particular case that is more frequently observed
in the Local Universe.

\begin{figure*}[!t]
  \includegraphics[angle=270,width=0.9\linewidth]{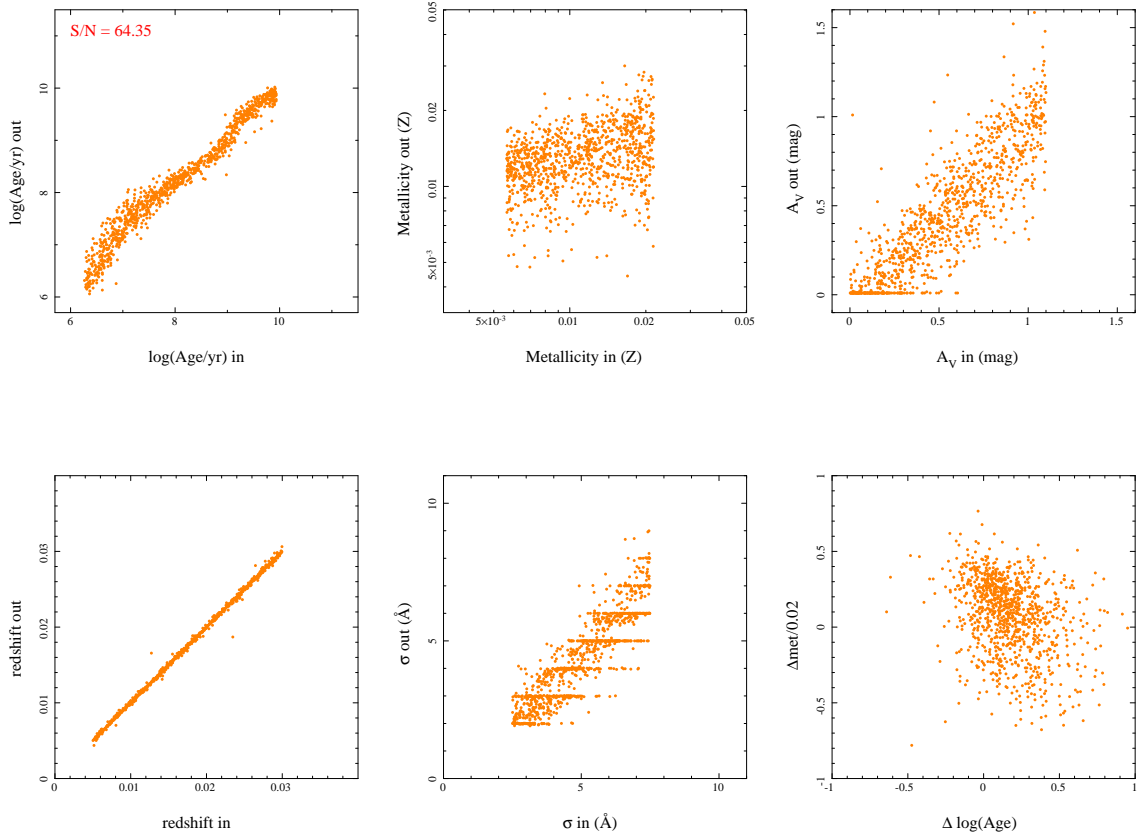}
  \caption{Results from the simulations. From left to right and from top to bottom, each panel shows the
    comparison between the input luminosity-weighted age, metallicity, dust attenuation,
    redshift, and velocity dispersion with the  values recovered by our fitting technique.
    In addition, the degeneracy between different parameters is explored, showing a comparison
    of the differences between the input and output values for age and metallicity,
     age and velocity dispersion, metallicity and velocity dispersion, and age
    and dust attenuation. The results are presented
    for the {\tt gsd156} template and the simulations for a S/N$\sim$60.}
  \label{fig:sim1}
\end{figure*}

In the second set of simulations we try to create a more realistic
SFH, in which the contribution of a stellar population within the
template is proportional to a power of its age in Gyr ($a_j \propto
Age_j^{1/2}$).  This is basically an exponential star formation rate
along the evolution of the stellar population. The metallicity distribution 
 follows a Gaussian probability function similar to the one
described before. However, in this particular case 
the input metallicity is forced to follow a linear anti-correlation with 
age; i.e.,  older stellar populations are more metal poor than 
younger ones, following the recent results by \citet{rosa14}.

\subsubsection{Results from the single burst simulation}
\label{rBurst}
%\label{ac_ssp}
%\label{ssp_lib}

For the first {\it single burst} SFH we create four different sets of
simulations. In each one, 1000 simulated spectra were created with the
same normalized flux at 5000 \AA, corresponding to 100, 50, 10 and 5
\funits, respectively.  The velocity was randomly selected
within the redshift range of the CALIFA footprint 
($0.005<z<0.03$), and the velocity dispersion was selected randomly
from $\sigma=2.5-7.5$ \AA\ ($\sigma_v\sim100-400$ km/s).
Then, the simulated spectrum was reddened by a dust attenuation randomly
selected between $A_V=0-1.1$ mag. Finally,  a purely poissonian
noise was added, to a level of 1 \funits. As indicated before, this is not a totally
realistic noise pattern. However, to perform a more detailed simulation
it would require to adopt {\it ad hoc} noise patterns, attached to a particular
data set \citep[e.g.][]{sanchez11}. 

The fitting procedure was then applied to the simulated spectra using
the same steps adopted to analyze the observational data, deriving,
for each one, the different data products described in
Sec. \ref{sec:average}. Figure \ref{fig:Age_Met_burst} illustrates the
result of this exercise for a particular spectrum with a
signal-to-noise ratio of $\sim$100.  It shows the input coefficients
for a particular simulated spectrum created using the {\tt miles72}
template library (for both the simulation and the analysis), in comparison
with the output distribution. This kind of distribution comprises
the star-formation history and metal enrichment pattern, that are
projections of this distribution. It is clear that the luminosity
weighted values derived from both distributions are very
similar { (log(age/yr)$_{\rm in}$=9.76 vs. log(age/yr)$_{\rm out}$=9.78 and [Z/H]$_{\rm in}$=-1.68 vs. [Z/H]$_{\rm out}$=-1.56 ), despite the fact that there are clear fluctuations between the input and recovered weighted/contributions of each particular stellar component. The largest discrepancy is for the lower metallicity bin and the log(age/yr)=9.50 component, that differs about a 10\% each one.} Furthermore, the distribution of coefficients in the
age-metallicity grid is remarkable similar.  Similar analyzes were
performed by \citet{ocvrik2006}, \citet{koleva2009}, and
\citet{sanchez11}, although they cover a more reduced parameter space,
 and in general the kinematics and/or the dust attenuation
were fixed in their simulations.

Table 1 lists, for each set of simulated spectra, their average
signal-to-noise ratio and the difference between the input and the
recovered values for the following parameters: luminosity-weighted
age, metallicity, and dust attenuation, together with the velocity and
velocity dispersion, with their corresponding standard deviation. Here
we are measuring two different effects: (i) the systematic deviation
from the input parameters, characterized by the average offsets, {
  (i.e., the accuracy of the measurement)} and (ii) the {
  precision} in the measurement of the parameter (with or without a
systematic bias or offset), characterized by the standard deviation.
{ This corresponds to the expected the precision of each individual
  realization.  However, the error of the systematic deviation
  described before is $\sim$30 times smaller, taking into account that
  each row corresponds to 1000 independent simulations.}  Only in the
case that the average offset is clearly larger than this { error}
it is possible to claim that there is a systematic offset { for all
  the simulated dataset}. This is { more} evident for the spectra
with high S/N {, where the systematic effects dominate over the
  precision of the measurement. However, in most of the simulated
dataset there is a clear and well defined bias with similar trend
for each different template, despite of the estimated precision}. { Nevertheless}, if the
standard deviation is clearly larger than the derived offset, { the
  expected bias for an individual realization is too small compared
  with the systematic bias/offset and therefore is sub-dominant.}  This is
the case { for} most of the values in the table, and in particular
for the spectra with lower S/N.

There are large differences between the accuracy and { precision}
of the recovered parameters depending on the adopted stellar
templates. For those templates with higher spectral resolution, {\tt
  miles72} and {\tt gsd156}, the parameters are recovered with a
better { accuracy and } precision for the spectra with higher
signal-to-noise ratio.  In those cases the standard deviation { and} the
offsets between the input and output values are smaller,
irrespectively of the possible systematic offsets. Those offsets are
in general smaller than the standard deviations, and therefore they
are { sub-dominant for individual realizations, although in general they
are statistically significant}. 
%There are few cases for which the systematic
%offset could be real, like in the case of the recovery of the age for
%the {\tt gsd156} template at a S/N$\sim$60. However, in all the cases
%it is below a 1$\sigma$ significance level.

In summary, the simulations indicate that it is possible to recover
the properties of the stellar populations for { at least a} S/N of
$\sim$\,30-50 per pixel to derive accurate { and precise} results in this particular
case.  However, for those templates with coarser spectral resolution
the parameters are recovered equally well/bad for a wide range of
S/N. Figure \ref{fig:sim1} illustrates these results, showing for a
particular set of simulations the comparison between the input and
recovered parameters for each individual spectrum in the dataset.
{ In some cases, like in the age-age comparison, there are clear structures
that indicates a possible bias in the recovery of this parameter.}

\begin{figure*}[!t]
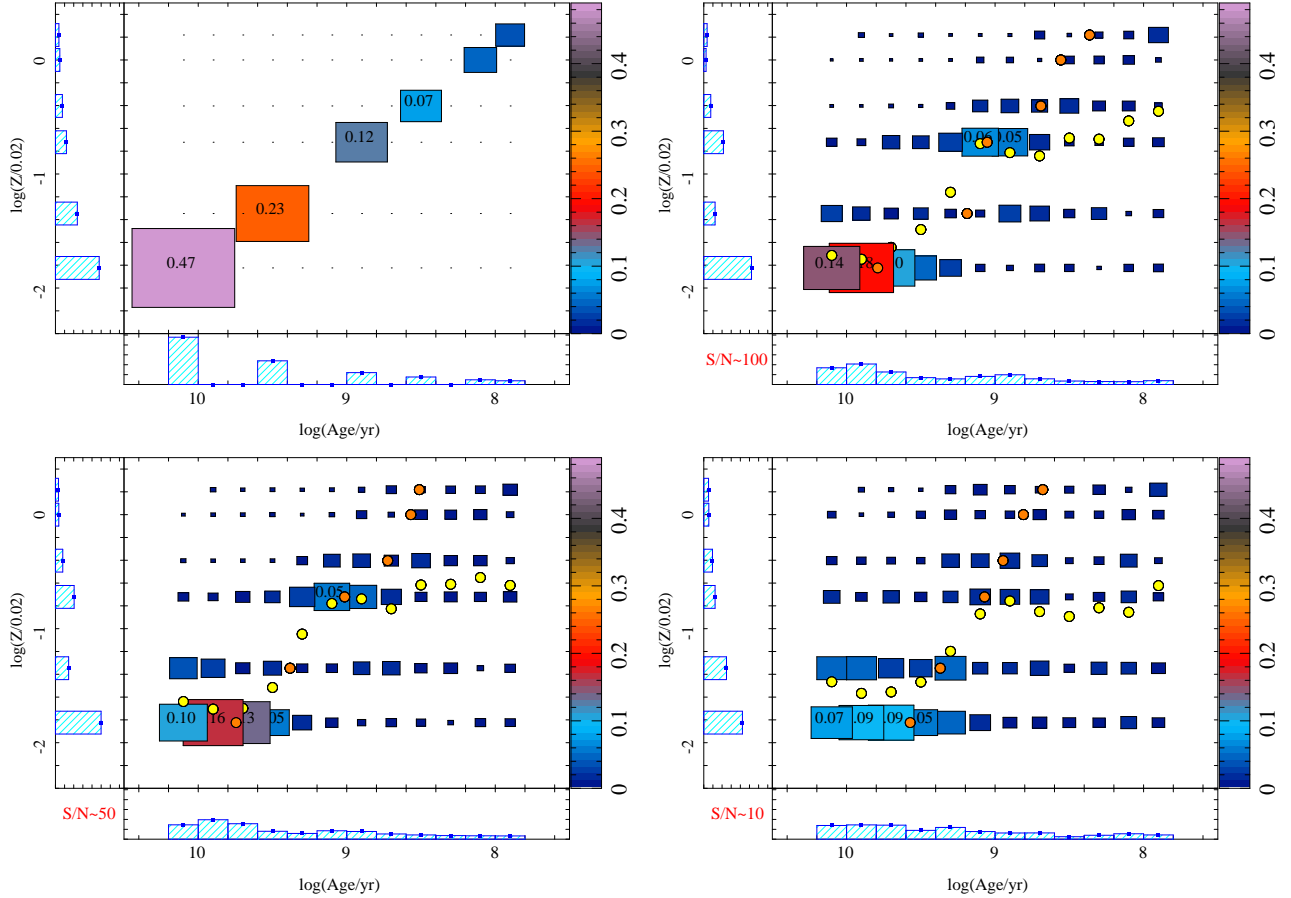

  \includegraphics[angle=270,width=0.51\linewidth]{figs/SFH_IN.ps}\includegraphics[angle=270,width=0.51\linewidth]{figs/SFH_OUT_100.ps}
  \includegraphics[angle=270,width=0.51\linewidth]{figs/SFH_OUT_50.ps}\includegraphics[angle=270,width=0.51\linewidth]{figs/SFH_OUT_10.ps}
  \caption{ {\it Top-Left panel}: Input coefficients distribution along the age-metallicity grid for a simulated spectrum corresponding to the second set of simulations, i.e, those resembling a star formation history with a continuous declining star formation rate, using the {\tt miles72} template library. For each age-metallicity pair the weight is shown as a solid square, with size and color  proportional to the weight of the corresponding SSP$_{age,met}$. For those SSPs with a stronger contribution, the actual fraction is indicated on top of each square. The horizontal and vertical histograms represent the co-added coefficients for a particular age and metallicity respectively. {\it Top-Right panel:} Similar distribution corresponding to the coefficients recovered for a spectrum with a signal-to-noise ratio of $\sim$100. The yellow filled circles show the average metallicity for a given age, while the orange filled circles show the average age for a given metallicity. {\it Bottom-left panel:} Similar distribution as shown in the previous panel but for a simulated spectrum with S/N$\sim$50. {\it Bottom-right panel:} Similar distribution as shown in the previous panel but for a simulated spectrum with S/N$\sim$10.}  \label{fig:Age_Met_SFH}
\end{figure*}

A comparison { of} the offset between the output and input values
for different parameters is also explored  to understand the possible
degeneracies among them. It particular, we explore the possible
bias introduced by the age-metallicity degeneracy: 
In general older stellar populations are redder, like metal rich ones, 
while younger and metal poor ones are bluer \citep{Worthey:1994p3434}. 
This introduces a degeneracy in the
derivation of both parameters, as discussed by previous authors
\citep[e.g.][]{patri11}. The strength of this degeneracy is
parametrized by the correlation coefficient, $r_{cor}$, between both parameters,
listed in Table 1. The correlation shows a similar behavior
for the different stellar templates when S/N>10. 
The offset in age and metallicity is in most cases of the order of $r_{cor}\sim$0.4-0.5.

The age-metallicity degeneracy shows the second strongest correlation. 
The weakest ones have not been included in the figure. 
They correspond to the possible
degeneracy between the metallicity and both the dust attenuation
and the velocity dispersion, that are of the order of
$r_{cor}\sim0.1$. Thus, there is no degeneracy between those
parameters.  The offset in age present a weak trend with the
offset in  velocity dispersion, with a typical
$r_{cor}\sim$0.3. Finally, the strongest correlation is found between
the offset of age and the offset of dust attenuation, for
which  the correlation is strong ($r_{cor}\sim$0.8) in all the cases explored. 
This degeneracy is a well known one that
indicates how important it is to have a good determination of the dust
attenuation to derive the ages. This degeneracy affects more to the
younger than to the older stellar populations.

It is beyond the goal of this study to make a comparison between the
properties of the different SSP libraries discussed here. We used
different ones to illustrate that the procedure is able to recover the
input parameters independently of a particular library. However, it is
clear that there are differences.  { This is the case of the
  already quoted {\tt miles72} and {\tt gsd156} templates, when
  compared with the {\tt bc138} one. The accuracy of the derived
  parameters is very similar in the three cases (as estimated by the
  bias/offset). However, the precision is much worst for the later
  one.}  By far, the parameters are recovered with less {
  precision} when using the libraries {\sc bc138} and {\tt
  mar136}. { This may reflects subtle effects of how the space of
  parameters is coveraged, differences in the stellar templates
  adopted, or details in the code that generate the templates.  For
  example, the {\tt miles72} template does not include SSPs younger
  than 60 Myr, while the {\tt gsd156} template does not include SSPs
  more metal poor than 0.2 $Z_\odot$, while the range of parameters
  covered by {\tt bc138} includes both of them. Finally, the {\tt
    bc138} template uses the STELIB stellar library \citep{stelib},
  which spectrophotometric calibration along the full wavelength range
  is worse than the one presented by the MILES one \citep{miles}.
  We have repeated analysis based on the {\tt bc138} library restricting
  its age/metallicity coverage to that of the {\tt miles72} and {\tt gsd156}.
  Only in the case of the metal-rich version of the library there is an 
  improvement on the accuracy { and precision} of the recovered parameters. However
  it never reach the values found for the two later libraries. Therefore, although
  the coverage in the age/metallicity space has an effect on the results of the
  simulations, biasing towards those libraries with lower metallicity range, it
  cannot explain totally the reported differences.}

{ Finally}, the {\tt mar136} library has the lowest spectral
resolution, that strongly affects the results. A detailed discussion
of the expected differences when using different stellar templates can
be found in \citep{rosa15a}. For the other libraries, the stellar
age, dust attenuation, and stellar velocity properties are recovered
well, although there is a wide range in the accuracy and { precision} of the properties
derived, as summarized in Table 1. For the SSP libraries with the
better spectral resolution, and for S/N$>$50, the { precision} is of the
order of $\sim$0.1-0.2 dex, similar to the one reported by other
fitting procedures \citep[e.g., {\sc starlight,
}][]{cid-fernandes14}. The reported uncertainties are slightly larger
than the ones estimated for other codes, like {\sc Steckmap}
\citep{ocvrik2006,patri11}.  \citet{ocvrik2006} found a typical error
$\Delta age=0.06-0.09$ dex and $\Delta met=0.10-0.14$ dex, for
signal-to-noise ratios of 50 and 20 respectively, and a spectral
resolution $R=1000$. However, in these latter simulations the effect
of the precision of the velocity dispersion and the dust attenuation
was not taken into account. Using a similar code, \citet{patri11}
found that there is an intrinsic uncertainty in the derivation of the
parameters associated with the velocity dispersion, a trend already
reported by \citet{koleva2009}.

The velocity dispersion is recovered better for those stellar
templates with higher spectral resolution (e.g., {\tt miles72}), and
worse for those with lower spectral resolution (e.g., {\sc
  mar136}). Unexpectedly, the velocity does not present the
same trend. There is an updated version of the \citet{mar05} SSP
models with higher spectral resolution, based on  different
empirical and theoretical spectral libraries \citep{mar11}. They cover
a similar range of ages as the  templates outlined here,
however, they do not cover the same range of metallicities for the
younger stellar ages. In general, they produce similar results than the
ones provided by the {\tt miles72} and {\tt gsd156} models.
Therefore, the main differences described here are introduced by the
coarse spectral resolution of the \citet{mar05} models. When the
updated/higher-resolution version is used, we find no significant
differences.

The age-metallicity degeneracy is weaker for the high signal-to-noise
simulations for the {\tt gsd156} and {\tt mar136} libraries, slightly
larger for  {\tt bc138}, and clearly stronger for  {\tt miles72}. 
However, in all  cases the correlation is significant, with a
probability of not being produced by a random distribution of the data
larger than a 99.99\%. In the case of the {\tt bc138} library the
same reason outlined before for the larger error in the 
parameters recovered could explain this small increase in the degeneracy. In
the case of  {\tt miles72}, the fact that this library does not
include stellar populations younger than 63 Myr could produce that
effect, since the algorithm tries to compensate the absence of these
young stellar populations giving more weight to more metal poor ones.
Finally, we notice that for some templates the strength of the
degeneracy decreases with increasing noise, which indicates that for 
noisy spectra the random error dominates over this systematic bias (e.g., {\tt gsd156} 
and {\tt miles72}). However, for other templates, the strength is enhanced by
the noise (e.g. {\tt mar136}), highlighting the effect of a coarse
spectral resolution in the degeneracy between both parameters.

A possible source of error in the interpretation of these simulations
is the fact that the same set of templates was used to create the
simulated data and to model them. In principle, we assumed that the
grid of templates is representative of the real stellar population of
the spectra analyzed, but it is likely that this library is
incomplete. This is not the case in our simulations, by
construction. To study the effects of this possible incomplete
representation of the observational spectra by the template library,
we created a new set of simulated data using the {\tt gsd156} library,
that was then fitted using a much reduced version of the template
library, that we called {\sc gsd32}. This template comprises a grid of
32 SSPs, corresponding to 8 ages covering the same range than {\sc
  gsd156}, and with the same 4 metallicities. The results from this
simulation are listed in Table 1. It shows that in the case
of an incomplete sampling of the spectroscopic parameters by the model
template, these parameters are still well reconstructed, albeit with a
lower { precision} (as expected).  We repeated the experiment with an even
more reduce set of stellar libraries, with only 12 SSPs, with a
similar result. It is important to note that in both cases the
more extreme stellar populations from the original template were retained in the
restricted one (i.e., the oldest/youngest and more metal rich/poor). 
Finally, we want to remark that although the average parameters that
characterize the stellar populations are still recovered well, the
details of the SFH and chemical enrichment are lost, obviously.

\subsubsection{Results from the simulated SFH}
\label{rSFH}
%\label{rBurst}
%\label{ac_ssp}
%\label{ssp_lib}

The second set of simulations is created to follow a more realistic
SFH. We create three different sets of
simulations, each one comprising 1000 simulated spectra with the same
normalized flux at 5000 \AA, corresponding to 100, 50 and 10 \funits,
respectively. The velocity, velocity dispersion, and dust
attenuation were taken into account in a similar way as in the previous set
of simulations described in Sec. \ref{rBurst}. For this study we
show only the results using the {\tt miles72} template library,
although we have repeated all the analysis using the other
libraries with similar results.

The fitting procedure was then applied to the simulated spectra using
the same steps adopted to analyze the observational data, deriving,
for each one, the different data products described in
Sec. \ref{sec:average}. Figure \ref{fig:Age_Met_SFH} illustrates the
result of this exercise for three particular spectra, each one
corresponding to a different S/N level.  It shows the input coefficients for a
particular simulated spectrum in comparison with the output
distribution. It is clear that both the SFH, represented by
the histogram in the bottom of each panel, is well reproduced for the
different simulated spectra, although as expected, the input shape is
more and more diluted as the S/N decreases. The metal enrichment
pattern is well reproduced in any of the simulations, being more
accurate for the older stellar populations than for the younger stellar populations. The distribution of
metallicities at different ages and ages at different metallicities,
represented in the figure as yellow and orange solid circles, are a
relatively good representation of the input correlation between the
metals at different ages; thus, they are good tracers of the metal
enrichment pattern.

%----------------------------------------------------------------
\begin{table}[t]
\begin{center}
\label{simtab_sfh}
\caption[Results of the simulations: simulated SFH]{Results of the simulations: simulated SFH.}
\begin{tabular}{rrrrcrr}\hline\hline
\multicolumn{1}{c}{S/N} & \multicolumn{1}{c}{$\Delta$Age/Age} & \multicolumn{1}{c}{$\Delta$Z/Z} \\\hline
%\tableline
%1368.9 & -0.12\,$\pm$0.01 & 0.00\, $\pm$0.01 \\
% 878.3 & -0.12\,$\pm$0.01 & 0.00\, $\pm$0.01 \\
 118.0 & -0.12\,$\pm$0.02 & 0.01\, $\pm$0.03 \\
  59.7 & -0.14\,$\pm$0.03 & 0.03\, $\pm$0.04 \\
  11.9 & -0.20\,$\pm$0.10 & 0.08\, $\pm$0.15 \\
\hline
\end{tabular}
\end{center}
\end{table}
%----------------------------------------------------------------

Table 2 lists, for each set of simulated spectra, their average
signal-to-noise ratio and the difference between the input and the
recovered values for two of the parameters derived:
luminosity-weighted age and metallicity. The kinematics and dust
attenuation values are not listed since they do not present
significant differences with the previous simulations, due to the
decoupling of the derivation of these properties and the derivation of
the SFH, as described in Sec. \ref{fitting}. We note that the { precision}
of the parameters recovered is better for this more realistic SFH than
for the single-burst one, described in Sec. \ref{rBurst}, even for low
S/N spectra.
%This is somehow expected, since the
%introduced anti-correlation between the age and the metallicity reduces
%the degree of freedom in the linear fitting regression. It is also
%very reassuring, since it is expected that most of real spectra of
%galaxies present a similar anti-correlation.
In general, the two luminosity-weighted parameters are recovered with
a { precision} better or of the order of 0.1 dex.

\begin{figure*}[!t]
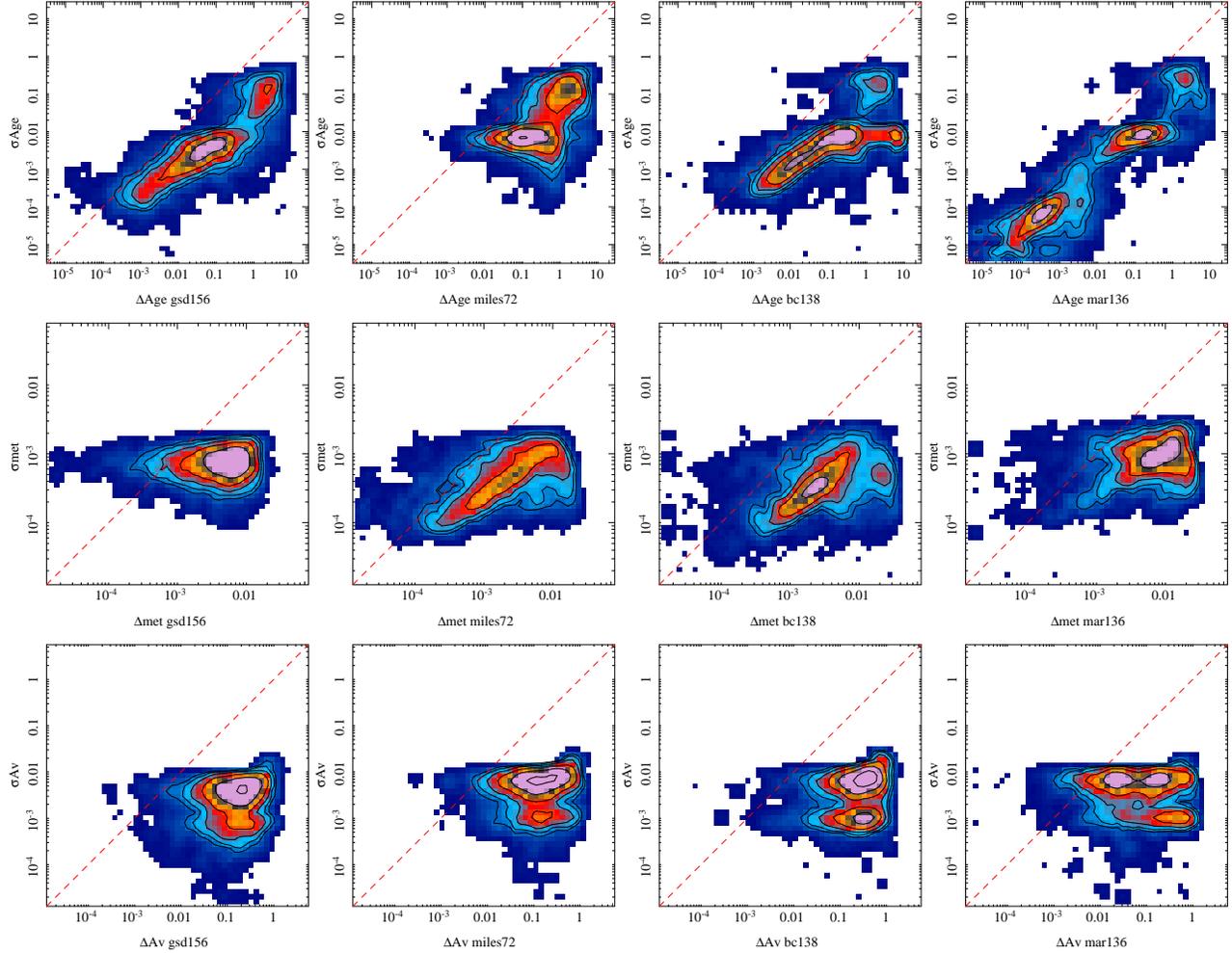

\includegraphics[angle=270,width=0.25\linewidth]{figs/DAge_eAge_gsd156.ps}\includegraphics[angle=270,width=0.25\linewidth]{figs/DAge_eAge_miles72.ps}\includegraphics[angle=270,width=0.25\linewidth]{figs/DAge_eAge_bc138.ps}\includegraphics[angle=270,width=0.25\linewidth]{figs/DAge_eAge_mar136.ps}
\includegraphics[angle=270,width=0.25\linewidth]{figs/Dmet_emet_gsd156.ps}\includegraphics[angle=270,width=0.25\linewidth]{figs/Dmet_emet_miles72.ps}\includegraphics[angle=270,width=0.25\linewidth]{figs/Dmet_emet_bc138.ps}\includegraphics[angle=270,width=0.25\linewidth]{figs/Dmet_emet_mar136.ps}
\includegraphics[angle=270,width=0.25\linewidth]{figs/DAv_eAv_gsd156.ps}\includegraphics[angle=270,width=0.25\linewidth]{figs/DAv_eAv_miles72.ps}\includegraphics[angle=270,width=0.25\linewidth]{figs/DAv_eAv_bc138.ps}\includegraphics[angle=270,width=0.25\linewidth]{figs/DAv_eAv_mar136.ps}
  \caption{Distribution of the estimated errors for the different parameters derived for the stellar populations, luminosity-weighted age (top panels), metallicity (middle panels), and dust attenuation (bottom panels), along with the absolute difference between the output and input parameter for the first set of simulations described in Sec. \ref{rBurst}. Each panel, from left to right, corresponds to a different stellar template library  considered in the simulations. For each panel the contours correspond to the density of points, with the first contour encircling 95\% of the points, and each consecutive contour encircling 15\% less number of points.}  \label{fig:eSSP_Age}
\end{figure*}

%\includegraphics[angle=270,width=0.25\linewidth]{figs/Dmet_emet_miles72.ps}

%A possible source of error in the interpretation of these simulations
%is the fact that the same set of templates was used to create the

\subsubsection{Accuracy on the estimated errors}
\label{rSSP_ER}
%\label{rSFH}
%\label{rBurst}
%\label{ac_ssp}
%\label{ssp_lib}

The two main goals of the current modification of {\sc FIT3D}
with respect to previous versions, regarding the analysis of the
stellar populations are: (i) to provide with a reliable estimation of
the star formation history and the average properties describing the
stellar population, and (ii) to obtain a useful estimation of the
uncertainties of the  parameters derived. In previous sections we have
illustrated how well are recovered the parameters that describe the
underlying stellar population. In this section we explore how
representative are the estimated errors of the real uncertainties in
these parameters.  For doing so we used the first set of simulations,
described in Sec. \ref{rBurst}, that cover a wider range of
parameters, and we compare the errors estimated by the fitting
procedure ($\sigma{\rm PAR}$) with the absolute difference between the
recovered and the input parameters ($\Delta{\rm PAR}$), where ${\rm PAR}$
is the parameter considered: age, metallicity, or dust attenuation ($A_V$).

{\sc FIT3D} estimates the error of each parameter 
based on the individual values derived along the MC-chain of
realizations described in the previous sections. The best fitting model is derived
for each MC realization, and therefore, the
best set of parameters that describe the input
spectrum. It is generally assumed that the standard deviation of the
values derived for each parameter is a good representation of
the real errors ($\sigma{\rm PAR}$). Indeed, this is the basic assumption of any MC or
bootstrapping procedure aimed to estimate the errors in any analysis.
However, this assumption holds if the different parameters are
independent, and if the errors affect equally each of them. Therefore,
there is no guarantee that $\sigma{\rm PAR}$ is indeed a good representation
of the real uncertainty in the derivation of the  parameter.

Figure \ref{fig:eSSP_Age} shows the distribution of $\sigma{\rm PAR}$
along $\Delta{\rm PAR}$, for the three  parameters, and for
each of the stellar template libraries described before. The patterns
in the comparison between the estimated and real errors are very
similar for the different  stellar templates, with most of the
differences due to the  ranges covered in age and
metallicity. First, it seems that the errors derived by the fitting
procedure are a good trace of the real differences between the input
and recovered values, for both the luminosity-weighted age and
metallicity, with a global underestimation of the real errors by a
factor $\sim$4, $\Delta{\rm Age}\sim4\sigma{\rm Age}$, and $\Delta{\rm met}\sim4\sigma{\rm met}$. 
For those libraries with a smaller range of
age or metallicity the trend between both errors is less obvious,
as expected since the dynamical range of parameters is less well covered.

For the dust attenuation, there is also a bias between the recovered
and real error, but it is less obvious that the estimated error is a
good estimation of the { precision} of { how} this parameter { is recovered}. Indeed, in general
it seems that the dust attenuation is recovered with an average
uncertainty of $\sim$0.1-0.3 dex, without a clear correspondence with $\sigma{\rm A_{\rm V}}$.
{ However, we must recall that the error in the dust attenuation was not
estimated on the basis of a MC-scheme, and it corresponds to the value estimated
from the $\chi^2$ curve. In further versions of {\sc FIT3D} we will try to implement
a MC-scheme for this critical parameter too.}

In summary, the basic assumption that $\sigma{\rm PAR}$ is a good
representation of the real uncertainty does not hold. For the Age and
Metallicity the estimated errors are a factor 4 lower than the real
uncertainties, although there is a correlation between both of them
and therefore it is possible to derive the later ones form the former
ones in a simple way (in general). However, for the dust attenuation,
the errors estimated by {\sc FIT3D} are unrealiable and there is no
simple way to transform them into realistic ones.

%, independently of the rest of the parameters involved in 
%the simulation.

\begin{figure*}[!t]
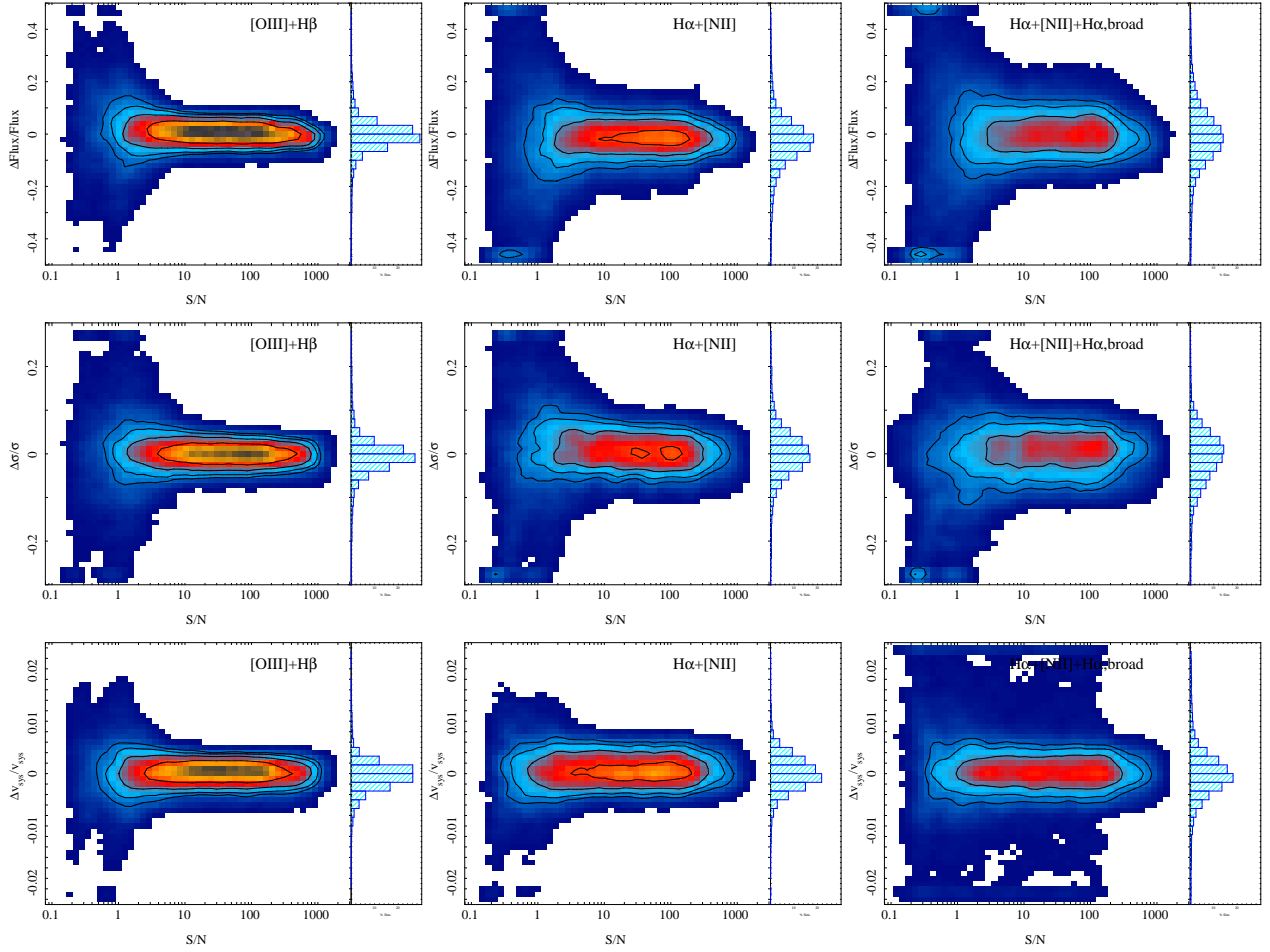

\includegraphics[angle=270,width=0.33\linewidth]{figs/DFlux_OIII_SN.ps}\includegraphics[angle=270,width=0.33\linewidth]{figs/DFlux_Ha_SN.ps}\includegraphics[angle=270,width=0.33\linewidth]{figs/DFlux_broad_SN.ps}
\includegraphics[angle=270,width=0.33\linewidth]{figs/Ddisp_OIII_SN.ps}\includegraphics[angle=270,width=0.33\linewidth]{figs/Ddisp_Ha_SN.ps}\includegraphics[angle=270,width=0.33\linewidth]{figs/Ddisp_broad_SN.ps}
\includegraphics[angle=270,width=0.33\linewidth]{figs/Dvel_OIII_SN.ps}\includegraphics[angle=270,width=0.33\linewidth]{figs/Dvel_Ha_SN.ps}\includegraphics[angle=270,width=0.33\linewidth]{figs/Dvel_broad_SN.ps}
  \caption{Relative difference between the recovered and the simulated parameters of the emission lines (top: flux intensity, middle: velocity dispersion, and bottom: velocity) for the three  sets of simulated emission lines corresponding to [\ion{O}{iii}]+H$\beta$ (left panels), [\ion{N}{ii}]+H$\alpha$ (central panels), and this later one with an additional broad component for H$\alpha$ (right panels), as a function of the signal-to-noise of the emission lines. For each panel the contours corresponds to the density of points, with the first contour encircling the 95\% of the points, and each consecutive contour encircling a 15\% less number of points.}  \label{fig:elines_SN}
\end{figure*}

\subsection{Accuracy of the properties of the emission lines}
\label{rEL}
%\label{rSSP_ER}
%\label{rSFH}
%\label{rBurst}
%\label{ac_ssp}
%\label{ssp_lib}

Following a similar philosophy, we create a set of
simulations in order to derive the accuracy { and precision} of the { recovery of the } properties
of the emission lines. The first set of simulations was created
assuming  purely poissonian noise, and without taking into
account the possible effects of the imperfect subtraction of the
underlying stellar population, and other sources of non poissonian
noise. We simulated three different systems of emission lines within
this set of simulations: (i) H$\beta$ and 
[\ion{O}{iii}]$\lambda$4959,5007, with a flux intensity of
100, 66, and 22 \funits, respectively, and covering the wavelength range
 4800-5050 \AA, in the rest-frame; (ii) H$\alpha$, 
[\ion{N}{ii}]$\lambda$6548,6583, and 
[\ion{S}{ii}]$\lambda$6717,6731, with a flux intensity of 100,
66, 22, 25, and 25 \funits, respectively,  covering a wavelength
range  6400-6900\AA; and (iii) the same emission lines
included in the previous simulation plus a broad H$\alpha$ component,
covering the same wavelength range.

All the emission lines, except the broad component included in
the last set, were simulated assuming the same range of velocity
dispersion used in the simulations of the stellar populations,
$\sigma=$2.5\AA\ and 7.5\AA\ ($\sigma_v\sim100-400$ km/s). The
velocity dispersion is then selected randomly  within this
range. Therefore, while in case (i) the emission lines are always
deblended, in case (ii) there are cases in which the emission lines
are deblended, but in many other cases they show different levels of
blending. Finally, in case (iii) the narrow emission lines present the
same degree of blending than in case (ii), but in all the cases they
are blended with the broad emission line. This  affects 
the accuracy { and precision} of the recovered emission line properties. For the broad
component included in case (iii) we consider a range of velocity
dispersion  $\sigma=32.5-37.5$ \AA\ ($\sigma_v\sim1500-1700$ km/s).

The velocity was randomly selected around $6000\pm500$ km/s, 
for the narrow emission lines. The broad
component is randomly redshifted by 200 km/s from this 
velocity, which is a typical situation in the case of broad components
generated by either outflows or type one AGNs. The emission line
spectra were sampled with a pixel scale of 2\AA/pix, similar to the
one provided by the CALIFA V500 data set. The results can be 
extrapolated to other instrumental configurations taking into account
the ratio between the velocity dispersion and the size of the sampling
pixel. The effect of the instrumental configuration has not been
taken into account. It affects mostly the recovering of the physical
velocity dispersion, which is a quadratic subtraction of the instrumental
one from the measured dispersion. Therefore, it basically depends on
this later parameter, although not in a linear way.
% We could simulate its
%effect {\it a posteriori}. 

%----------------------------------------------------------------
\begin{table}[!t]
\begin{center}
\label{simtab_e}
\caption[Results of the simulations: Emission lines]{Results of the simulations: Emission lines.}
\begin{tabular}{rrrr}\hline\hline
\multicolumn{4}{c}{Results for the [\ion{O}{iii}]+H$\beta$ emission lines} \\\hline
\multicolumn{1}{c}{S/N} & \multicolumn{1}{c}{$\Delta$Flux/Flux} &
\multicolumn{1}{c}{$\Delta$$\sigma_v$/$\sigma_v$} & \multicolumn{1}{c}{10$\Delta$v$_{sys}$/v$_{sys}$$^1$} 
\\\hline
$<$3     &     0.01\,$\pm$0.15 &    0.00\,$\pm$0.11 &    0.00\,$\pm$0.06 \\
3$-$10   &     0.00\,$\pm$0.05 &    0.00\,$\pm$0.04 &    0.00\,$\pm$0.03 \\
10$-$100 &     0.01\,$\pm$0.04 & $-$0.01\,$\pm$0.02 &    0.00\,$\pm$0.02 \\
$>$100   &  $-$0.01\,$\pm$0.03 & $-$0.01\,$\pm$0.02 &    0.00\,$\pm$0.02 \\
\hline
\multicolumn{4}{c}{Results for the [\ion{N}{ii}]+H$\alpha$ emission lines} \\\hline
\multicolumn{1}{c}{S/N} & \multicolumn{1}{c}{$\Delta$Flux/Flux} &
\multicolumn{1}{c}{$\Delta$$\sigma_v$/$\sigma_v$} & \multicolumn{1}{c}{10$\Delta$v$_{sys}$/v$_{sys}$} \\\hline
$<$3     &  $-$0.02\,$\pm$0.24 &    0.01\,$\pm$0.14 &    0.00\,$\pm$0.09 \\
3$-$10   &  $-$0.02\,$\pm$0.08 &    0.01\,$\pm$0.05 &    0.00\,$\pm$0.04 \\
10$-$100 &  $-$0.03\,$\pm$0.07 &    0.00\,$\pm$0.04 &    0.00\,$\pm$0.03 \\
$>$100   &  $-$0.03\,$\pm$0.05 & $-$0.01\,$\pm$0.04 &    0.00\,$\pm$0.03 \\
\hline
\multicolumn{4}{c}{Results for the [\ion{N}{ii}]+H$\alpha$+H$\alpha,broad$ emission lines} \\\hline
\multicolumn{1}{c}{S/N} & \multicolumn{1}{c}{$\Delta$Flux/Flux} &
\multicolumn{1}{c}{$\Delta$$\sigma_v$/$\sigma_v$} & \multicolumn{1}{c}{10$\Delta$v$_{sys}$/v$_{sys}$} \\\hline
$<$3     &  $-$0.01\,$\pm$0.17 & 0.00\,$\pm$0.09 &    0.00\,$\pm$0.08 \\
3$-$10   &  $-$0.01\,$\pm$0.10 & 0.00\,$\pm$0.06 &    0.00\,$\pm$0.08 \\
10$-$100 &  $-$0.01\,$\pm$0.09 & 0.00\,$\pm$0.05 &    0.00\,$\pm$0.07 \\
$>$100   &  $-$0.01\,$\pm$0.07 & 0.00\,$\pm$0.04 &    0.00\,$\pm$0.05 \\
\hline\end{tabular}
\end{center}

(1) $\Delta$v$_{sys}$ is multiplied by ten to show it at the same scale of the other
two parameters.

\end{table}
%----------------------------------------------------------------

Figure \ref{fig:elines_SN} shows the comparison between the recovered
and simulated parameters for three sets of simulated emission lines as
a function of the signal-to-noise ratio of the emission lines. The S/N
is estimated as the ratio between the peak intensity of the 
emission line and the noise level included in the simulated
spectrum. As expected there is a clear dependence with S/N (for
S/N$<$10), with the worst recovered parameters corresponding to the
lower S/N emission lines. For large S/N values ($>$20) there is almost
no dependence of the uncertainty with the S/N. There is also a clear difference between
the three sets of simulations, with the better { precision} corresponding
to case (i), and the worst to case (iii). This is not an effect of the
signal-to-noise, since in case (i) and case (ii) all the emission
lines have the same integrated fluxes and a similar flux density per
spectral pixel for a fixed dispersion.

A summary of these results is included in Table 3. For each set of
simulated emission lines, it lists the mean and standard deviation of
the relative difference between the recovered and simulated parameters
for different ranges of signal-to-noise ratio. The program is able to
recover the intensity of the emission lines with a { precision}
better than $\sim$10\% for weak emission lines (S/N$\sim$3-10), in all
the cases.  The { precision} becomes worse, $\sim$20\%, for
emission lines swimming in the noise, { although the accuracy is
  equally good}.  However, at this low S/N values the systematic
effects (in the subtraction of the underlying stellar population and
other sources of non Poissonian noise not considered in this
simulation) will dominate, and therefore this { precision} should be
considered as an upper limit. It is interesting to notice that
although the { precision} increases towards higher signal-to-noise
ratios, it never gets better than $\sim$3\%, { at least for the highest S/N ratios
explored in this simulations}. We consider that this threshold is related to the
fact that we fit together the intensity, velocity, and velocity
dispersion, and their derivation is somehow correlated. Therefore, the
S/N of the integrated flux alone may not be a good parameter to
characterize how well the emission lines are detected.

The velocity and velocity dispersion are also well
recovered.  In the case of the velocity dispersion it has a
similar { precision} as the emission line intensity; it is recovered with a
precision better than 10\% in almost all the simulated spectra. The
best { precision} for this parameter is $\sim$2\%, even for
high S/N spectra. Finally, the velocity is recovered with a
very high precision, $\sim$10 times better than
the other two parameters. The precision ranges $\sim$12--54
km/s at the average redshift of the simulated spectra. This precision
is between two and ten times better than the simulated velocity
dispersion. These simulations confirm a widely accepted result
indicating that the velocity derived using emission lines
could be recovered with a precision as good as a few per cent of the
velocity dispersion of the emission lines (typically 10\%).

As already noticed, the { precision} of the  emission lines parameters 
recovered is different for each set of simulations, being, in
general, worst for the simulations with blended emission lines.
The degradation of the { precision} could be as bad as two times the { precision}
if the emission lines are not blended.

\begin{figure*}[!t]
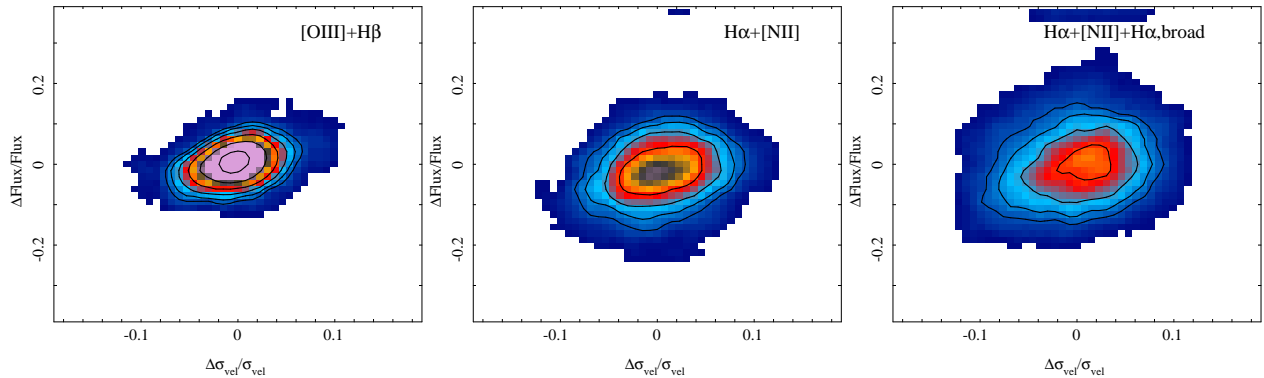

\includegraphics[angle=270,width=0.33\linewidth]{figs/DFlux_Dsigma_OIII.ps}\includegraphics[angle=270,width=0.33\linewidth]{figs/DFlux_Dsigma_Ha.ps}\includegraphics[angle=270,width=0.33\linewidth]{figs/DFlux_Dsigma_broad.ps}
  \caption{Relative difference between the recovered and the simulated  intensity of the emission lines for the three sets of simulated emission lines corresponding to [\ion{O}{iii}]+H$\beta$ (left panel), [\ion{N}{ii}]+H$\alpha$ (central panel), and this later one with an additional broad component for H$\alpha$ (right panels), as a function of the relative difference between the recovered and the simulated velocity dispersion of the corresponding emission lines. For each panel the contours correspond to the density of points, with the first contour encircling the 95\% of the points, and each consecutive contour encircling a 15\% less number of points.}  \label{fig:elines_disp}
\end{figure*}

\subsubsection{Interdependence of the accuracy between parameters}
\label{rEL_sigma}
%\label{rEL}
%\label{rSSP_ER}
%\label{rSFH}
%\label{rBurst}
%\label{ac_ssp}
%\label{ssp_lib}

As indicated before, the precision in the estimation of one of the
parameters may affect the accuracy with which others are recovered. In
particular, the velocity dispersion could affect the recovered flux 
intensity, since this parameter is an integral that depends on the
former one. Figure \ref{fig:elines_disp} shows the { relative} accuracy of the
recovered flux intensity as a function of the accuracy of the
recovered velocity dispersion for the three sets of simulations
described before. In all three cases the distributions show a cloud
of points with a dispersion of the order of
$\pm$0.1$\Delta_{par}/par$.  However, it is clear that the
distribution is broader and more roundish for the simulations
including blended emission lines than for the simulations including
only well resolved emission lines (left panel). This is a consequence
of the larger errors/inaccuracies in the derivation of both the flux
intensity and the velocity dispersion for the blended emission lines,
as shown in Section \ref{rEL}. { In this analysis we used all the simulations
described before irrespectively of the S/N. We repeated the analysis for different
S/N ranges without significant differences. Only for the very low S/N ($<$10) regime
we found a broadening of the distribution, but the main trends hold.}

There is a clear trend between both relative differences, that are
stronger for the simulations with un-blended emission lines, with a
correlation coefficient $r=$0.41 ($p>99.99\%$ of not being produced
by a random distribution of the data). This indicates that the bias
introduced in the estimation of the flux intensity by the inaccuracy
in the derivation of the emission lines is stronger for clearly
blended emission lines. However it is not a dominant factor
in the error budget, in the signal-to-noise range of our simulations.

\begin{figure*}[!t]
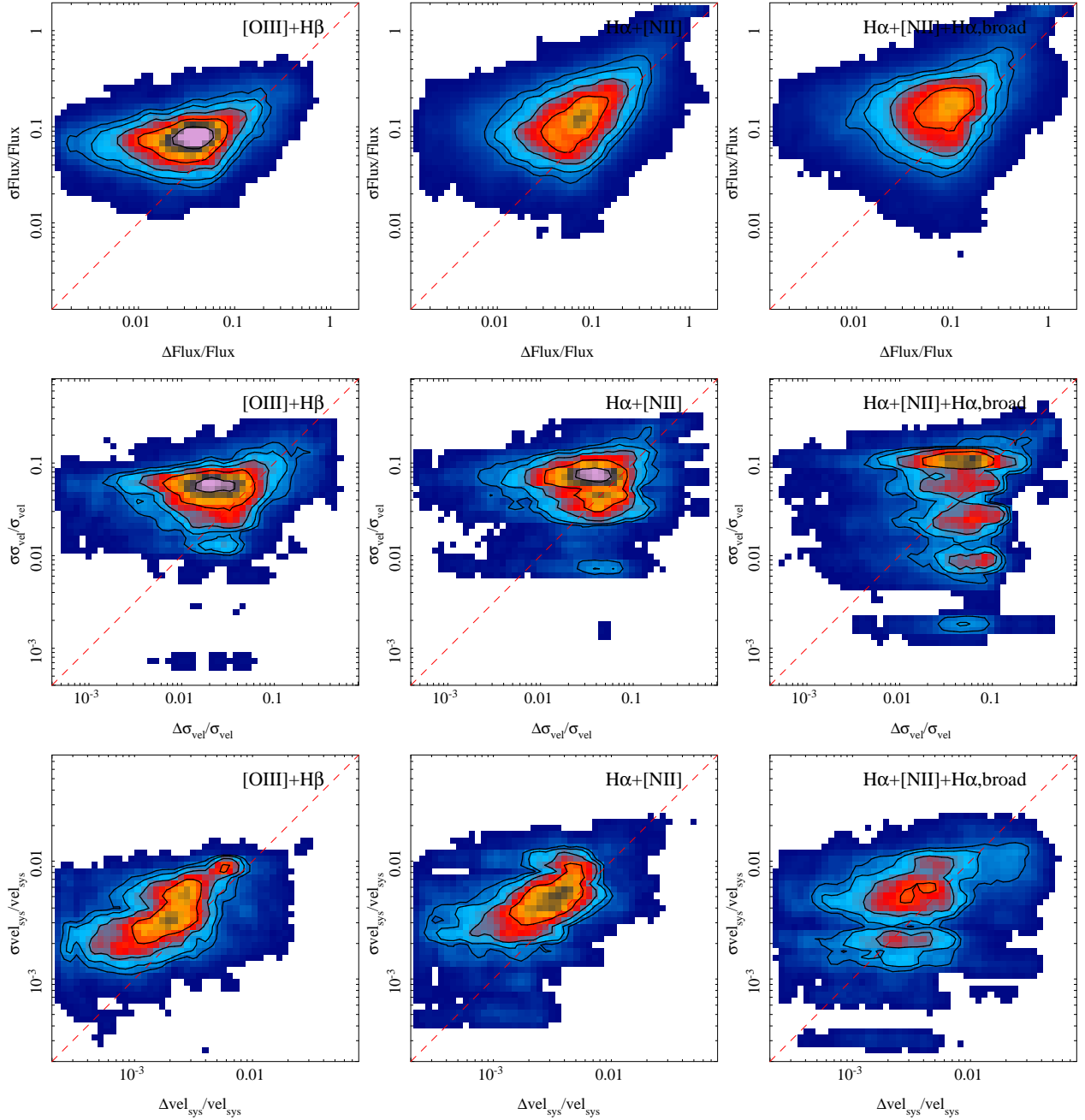

\includegraphics[angle=270,width=0.33\linewidth]{figs/DFlux_eFlux_OIII.ps}\includegraphics[angle=270,width=0.33\linewidth]{figs/DFlux_eFlux_Ha.ps}\includegraphics[angle=270,width=0.33\linewidth]{figs/DFlux_eFlux_broad.ps}
\includegraphics[angle=270,width=0.33\linewidth]{figs/Ddisp_edisp_OIII.ps}\includegraphics[angle=270,width=0.33\linewidth]{figs/Ddisp_edisp_Ha.ps}\includegraphics[angle=270,width=0.33\linewidth]{figs/Ddisp_edisp_broad.ps}
\includegraphics[angle=270,width=0.33\linewidth]{figs/Dvel_evel_OIII.ps}\includegraphics[angle=270,width=0.33\linewidth]{figs/Dvel_evel_Ha.ps}\includegraphics[angle=270,width=0.33\linewidth]{figs/Dvel_evel_broad.ps}
  \caption{Relative error estimated for the emission lines recovered parameters (top: flux intensity, middle: velocity dispersion, and bottom: velocity) for the three sets of simulated emission lines corresponding to [\ion{O}{iii}]+H$\beta$ (left panel), [\ion{N}{ii}]+H$\alpha$ (central panel), and this later one with an additional broad component for H$\alpha$ (right panels), as a function of the relative difference between the recovered and the simulated value of this parameter for the corresponding emission line. For each panel the contours correspond to the density of points, with the first contour encircling 95\% of the points, and each consecutive contour encircling a 15\% less number of points. The red dashed line in each panel corresponds to the one-to-one relation.}  \label{fig:elines_error}
\end{figure*}

\subsubsection{Accuracy on the estimated errors for the emission lines}
\label{rEL_ER}
%\label{rEL_sigma}
%\label{rEL}
%\label{rSSP_ER}
%\label{rSFH}
%\label{rBurst}
%\label{ac_ssp}
%\label{ssp_lib}

As in the analysis of the stellar population, it is
important to determine whether the estimated errors are representative of
the real ones. Following the previous analysis, we compare for each
emission line in each of the three simulated datasets the estimated
error with the real ones, i.e., the absolute difference between the
value recovered   and the simulated
(input) one. Figure \ref{fig:elines_error} shows the results of this
comparison.

As  shown in previous sections, in general, the real error in the
derivation of the parameters is larger for the simulations including
blended emission lines than in those including only un-blended ones.
For the flux intensity, the estimated errors seem to be a good
representation of the real ones for {\it real} errors larger than 
$\sim$4-5\%. For errors smaller than this value the program seems to
overestimate the errors, and derive a minimum value of $\sim$5\%,
regardless of the real difference between the input and output
parameters. However, since in most simulations it seems that
the emission intensities cannot be recovered with a { precision} better
than a few percent, this lower-limit still seems reasonable.

For the velocity dispersion the estimated errors seem to be of the
same order of the real ones, at least for the simulations that do not
include broad emission lines. In general, the errors seem to be
slightly overestimated, with a similar trend towards a minimum value
of the error around $\sim$5\%. For the simulations including broad
emission lines the estimation of the error shows a wide scatter,
with a clear underestimation of the error in many cases. We have
investigated in detail, and the larger variations in the estimation of
the error correspond to the broad emission line component included in
this simulation. It is interesting to note that, in any case, the
dispersion seems to be well recovered (see Fig. \ref{fig:elines_SN}),
but for broad emission lines the program tends to underestimate the
errors.

Finally, the estimated errors for the velocities are an order of
magnitude lower than  the errors derived for the other two parameters.
They are just slightly overestimated compared to the real difference between
the input and recovered velocities, being a good representation of the
real errors in general. As in previous cases, the estimated errors are
more accurate for the simulations including only narrow emission lines.

\begin{figure*}[!t]
  \includegraphics[angle=270,width=0.9\linewidth]{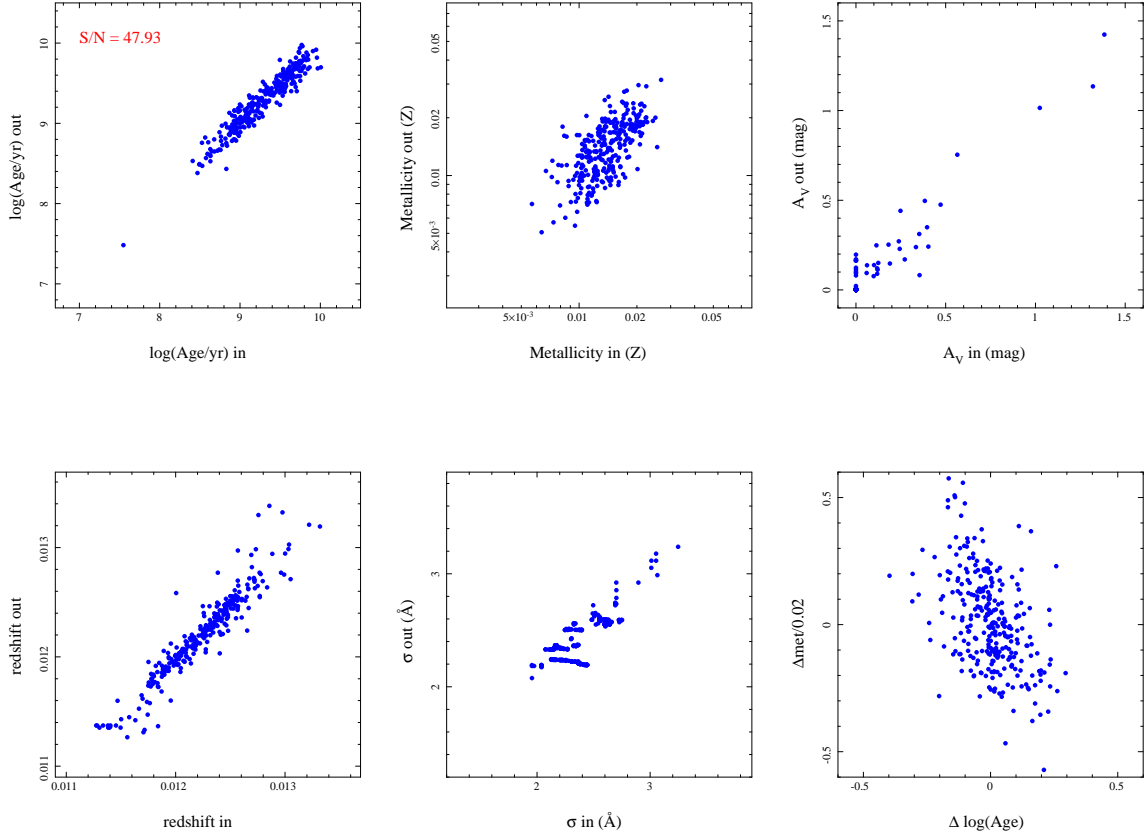}
  \caption{Results from the simulations based on the CALIFA V500 data cube of NGC\,2916. 
  Each panel shows, from left to right and from top to bottom, the comparison between 
  the input luminosity-weighted age, metallicity, dust attenuation, redshift, and velocity dispersion 
  with the values recovered by our fitting technique. The results are presented for {\tt gsd156} template. 
  The final bottom-right panel illustrates the degeneracy between the recovered age and metallicity, 
  showing a comparison between the input-output differences for both parameters.}
  \label{fig:ssp_NGC2916}
\end{figure*}

\subsection{Simulations using real data}
\label{simReal}
%\label{rEL_ER}
%\label{rEL_sigma}
%\label{rEL}
%\label{rSSP_ER}
%\label{rSFH}
%\label{rBurst}
%\label{ac_ssp}
%\label{ssp_lib}

So far we have estimated the accuracy { and precision} of the  parameters using
simulations that only take into account the Poissonian noise. 
To investigate further the precision in the derivation of the parameters
we require to take into account  all the different sources of error 
that are included in real data.

For these particular simulations we use the data sets provided by the
CALIFA survey. In particular, we adopted the V500 data cube
corresponding to NGC 2916, an Sbc galaxy at $z\sim$0.01244, that is
an average galaxy in this survey. This data cube comprises
78$\times$72 individual spectra, of which $\sim$2500 contain data and
the rest are out of the original hexagonal pattern of the instrument.
We need to remark that the simulations do not depend on the particular
galaxy selected. We used this one since it presents both strong and
weak emission lines within the FoV of the data cube, and therefore they
cover a wide range of signal-to-noise ratios. 

The simulations are created using as input the results from the
fitting using {\sc Pipe3D} over this particular dataset. The analysis
with {\sc Pipe3D} will be detailed in forthcoming articles of this
series, although some examples of the data products have been presented
already elsewhere \citep{sanchez13}. For the purpose of this
study we should know that the pipeline performs a spatial binning
over the data cube, aggregating adjacent spaxels based on their
difference in relative intensity (at a reference wavelength); 
it provides a final row-stacked spectra (RSS) comprising $\sim$500-1000 individual spectra
with an increased signal-to-noise (in the continuum) with respect to the individual
spaxels. Finally, each of the individual spectra
is analyzed using the current version of FIT3D to provide all the
data products regarding the stellar population and the emission lines
described in previous sections. In this particular analysis we adopted
the results derived using the {\tt gsd156} stellar library template.

%----------------------------------------------------------------
\begin{table*}[!th]
\begin{center}
\label{simtab_NGC2916}
\caption[Results of the simulations: simulated NGC\,2916 data cube]{Results of the simulations: simulated NGC\,2916 data cube.}
\begin{tabular}{rrrrcrr}\hline\hline
%\multicolumn{6}{c}{Results using the {\tt gsd156} SSP template} \\\hline
\multicolumn{1}{c}{S/N} & \multicolumn{1}{c}{$\Delta$Age/Age} &
\multicolumn{1}{c}{$\Delta$Z/Z} & \multicolumn{1}{c}{$\Delta$A$_V$$^1$} &
\multicolumn{1}{c}{$\Delta$v$_{sys}$$^1$} & \multicolumn{1}{c}{$\Delta$$\sigma_v$$^1$} 
& \multicolumn{1}{c}{$r_{cor}$$^2$} 
\\\hline
%\tableline
47.9 &  0.00\,$\pm$0.11 & 0.00\, $\pm$0.12 &  0.01\,$\pm$0.08 & -10.8\,$\pm$ 40.7 &  2.7\,$\pm$ 7.3 & 0.47\\
19.5 & -0.06\,$\pm$0.26 &-0.01\, $\pm$0.14 &  0.07\,$\pm$0.28 & -57.3\,$\pm$144.7 &  4.6\,$\pm$ 7.8 & 0.08\\
11.8 & -0.09\,$\pm$0.37 &-0.04\, $\pm$0.14 &  0.14\,$\pm$0.48 & -57.4\,$\pm$166.1 &  6.6\,$\pm$ 3.3 & 0.35\\
 5.4 & -0.31\,$\pm$0.38 &-0.03\, $\pm$0.11 &  0.66\,$\pm$0.58 &-186.0\,$\pm$218.5 &  7.4\,$\pm$ 7.4 & 0.26\\\hline
\end{tabular}
\end{center}
\end{table*}

%with a flux intensity, velocity and velocity dispersion maps for the
%stronger emission lines within the FoV and wavelength range:
%[\ion{O}{ii}]$\lambda$3727, H$\beta$,
%[\ion{O}{iii}]$\lambda$4959,5007, [\ion{N}{ii}]$\lambda$6548,6583,
%H$\alpha$ and [\ion{N}{ii}]$\lambda$6717,6731. For the weaker emission lines,
%it provides with a flux intensity derived adopting the kinematics properties
%of H$\alpha$, and extracting the integrated flux by coadding the intensity
%weighted by the assumed Gaussian shape. Thus, for the weak emission lines
%it is not performed a fitting procedure, but a Gaussian extraction.

The resulting star formation history, stellar kinematics, stellar dust
attenuation, and the corresponding data products of the emission lines
(intensity, velocity, and velocity dispersion) can be used as
input parameters for a simulation with more realistic parameters. For
each spatial bin we create a spectrum with a model stellar continuum
and a model set of emission lines, using as input the output of
the fit over the real data. Then, we renormalize the flux in each
individual spaxel within each bin following the relative contribution
of the flux to the bin by this spaxel in the original data. In
this way we have a realistic model data cube with the same spatial
distribution of the stellar population and ionized gas as the real data.

Then, we use the residuals from the fitting process over the observed
data cube to generate a realistic distribution of the noise, that 
comprises not only the Poissonian noise, but any other source of
error. To do so, we use the residual data cube obtained from the
analysis with FIT3D, and, for each spaxel at each wavelength, we
coadd a random value within a range plus-minus the median of the
absolute value of the noise distribution within a box of
$\pm$3$\arcsec$. In this way we respect the amplitude of the real
noise at each wavelength and at each position within the data cube,
excluding the peaks in the noise distribution (like the ones
produced by cosmic-rays or inaccuracies in the sky subtraction).
We should clarify that this derivation of the noise does not follow
either a Gaussian or a square distribution.

We then  apply the same spatial segmentation to extract the RSS-spectra
at the same spatial locations than in the case of the original
data cube. After that, we process the spectra using the same fitting
routines to recover both the properties of the stellar population and
emission lines. We adopted the same stellar population library for
the analysis, for a better comparison between the input and
output parameters. We analyzed four sets of emission lines, with
coupled kinematics, as described in Sec. \ref{sec:elines}. Those
emission lines are analyzed in four wavelength ranges: (1)
[\ion{O}{ii}]$\lambda$3727 (3700-3750\AA); (2) H$\beta$ and 
[\ion{O}{iii}]$\lambda$4959,5007 (4800-5050\AA); (3) H$\alpha$
and  [\ion{N}{ii}]$\lambda$6548,6583  (6680-6770); and (4)
 [\ion{S}{ii}]$\lambda$6717,6731.  Altogether, they comprise
more than 5000 individual emission lines, with different line intensities,
at a wide range of wavelengths. Therefore, they are affected in a different way by
the imperfect subtraction of the underlying stellar population and the
wavelength dependence depth of the data. All the emission lines
considered here are deblended at the spectral resolution of the
V500 CALIFA data, however, some of them could be affected by the wing of the
adjacent emission lines, in particular the doublet [\ion{S}{ii}]$\lambda$6717,6731.

\begin{figure}[!t]
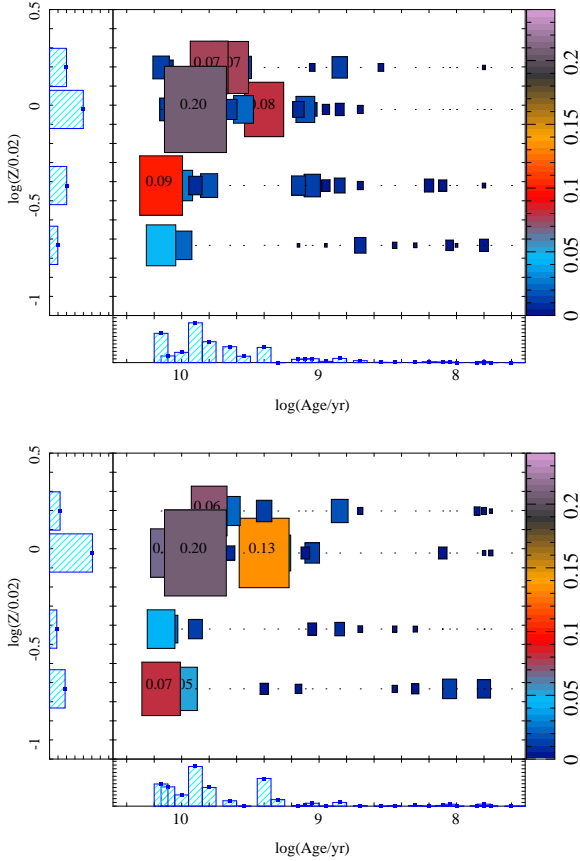

\includegraphics[angle=270,width=1.0\linewidth]{figs/SFH_NGC2916_in.ps}
\includegraphics[angle=270,width=1.0\linewidth]{figs/SFH_NGC2916_out.ps}
  \caption{ {\it Top panel}: Coefficient distribution along the age-metallicity grid derived from the analysis using {\sc FIT3D} of the central 5$\arcsec$ diameter aperture spectrum extracted from the CALIFA V500 data cube of NGC\,2916, used as input coefficients in the simulated spectra described in the text.  {\it Bottom panel:} Similar distribution corresponding to the recovered coefficients for the simulated spectrum, resulting from the fitting procedure. All the symbols are similar to those described in Fig. \ref{fig:Age_Met_SFH}.}  \label{fig:Age_Met_NGC2916}
\end{figure}

\subsubsection{Accuracy of the properties recovered for the stellar populations}
\label{simReal_SSP}

Figure \ref{fig:ssp_NGC2916} shows the results of  comparing
the  data products derived for the stellar populations as a function of
the corresponding input/simulated values, for  spectra with S/N
higher than 20. As in the case of Fig. \ref{fig:sim1}, it
demonstrates that there is a good correspondence between the input
and the recovered parameters, with a clear one-to-one correlation
in all the cases. A more quantitative comparison is included in
Table 4, that shows the average S/N and the difference between the input and the
recovered values, for different ranges of S/N. 
Despite the different procedure adopted, 
the results are very similar to the ones shown in Tables
1 and 2. For spectra with a S/N higher
than $\sim$20, most  parameters are recovered with a { precision of}
$\sim$10\%.  In the case of the velocity it is an order of magnitude 
better. It is interesting to note
that the velocity is recovered with a better precision when
using a more realistic SFH than a purely simulated one. 

%Maybe this is
%due to the well known radial gradients of the different 
%properties in galaxies, that impose secondary correlations between
%the parameters.  Due to this, they are not independent parameters, and
%the derivation of one of them affects the derivation
%of the others. This seems to be particularly important for the
%velocity dispersion, since we did not impose any correlation between
%this parameter and the ages and metallicities of the stellar
%populations in the simulations described in Sec. \ref{rBurst} and
%\ref{rSFH}.

\begin{figure*}[!t]
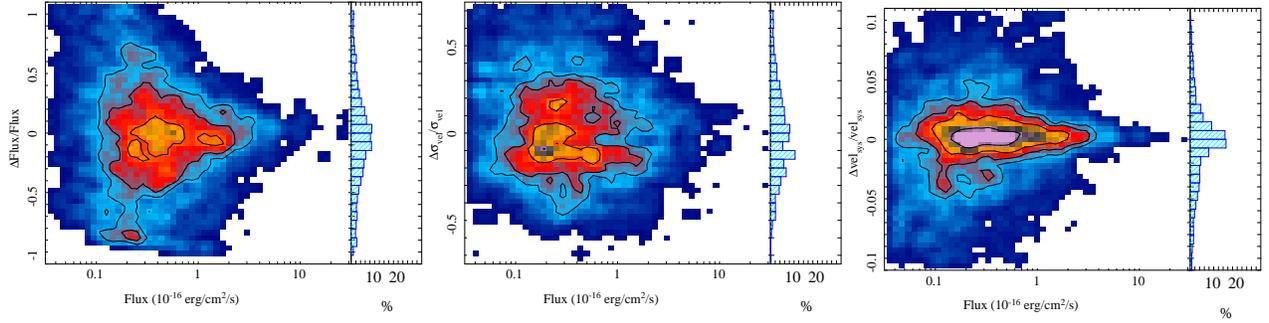

\includegraphics[angle=270,width=0.33\linewidth]{figs/DFlux_Flux_NGC2916.ps}\includegraphics[angle=270,width=0.33\linewidth]{figs/Ddisp_Flux_NGC2916.ps}\includegraphics[angle=270,width=0.33\linewidth]{figs/Dvel_Flux_NGC2916.ps}
  \caption{Relative difference between the recovered and input values for the parameter analyzed for the emission lines (left: flux intensity, center: velocity dispersion and right: velocity) for the simulation based on the CALIFA V500 data of NGC\,2916 as a function of the emission lines flux intensities. For each panel the colored images and contours correspond to the density of points, with the first contour encircling 95\% of the points, and each consecutive contour encircling a 15\% less number of points.}  \label{fig:lines_NGC2916}
\end{figure*}

As in the previous cases, we explore the possible
effect of the age-metallicity degeneracy in these more realistic
simulations.  The last panel of Fig. \ref{fig:ssp_NGC2916} shows a
comparison of the offset between the output and input ages, and
the offset between the output and input metallicities of the stellar
populations. There is a weak trend between the two offsets, in the
expected direction induced by this degeneracy.  The strength
of this degeneracy is parametrized by the correlation coefficient
between the two parameters, listed in Table 4.  The
trend is weak, and affects relatively more to  
metallicity than to age. This is expected since the relative error
in the derivation of age is always lower than the error in 
the derivation of metallicity. As expected, this
degeneracy, that is a systematic effect, is enhanced by the
S/N of the data. The effect is stronger for the more metal
poor stellar populations, since the offset ranges between $\pm$0.5 Z/Z$\odot$,
i.e., $\Delta Z =$$\pm$0.01. For the metal rich its effect is marginal, but
for the stellar populations with sub-solar abundances it could bias the result
by a factor 2 or more.

We note here that the degeneracy and the range of offsets between the
input and output parameters is larger than the values reported in
previous studies \citep[e.g.][]{ocvrik2006,patri11}, as we indicated
before.  Although they are of same order as the ones reported by more
recent simulations \citep[e.g.][]{cid-fernandes14}. Maybe the reason
for this disagreement is that, although in the previous cases 
 a full SFH was considered and reasonable chemical enrichment processes,
they did not explore a wide range of possible uncertainties in the
dust attenuation and the velocity dispersion, although they have a
hint of its effect \citep[e.g., Fig. B3 of ][]{patri11}.

%However, it is important to note
%that all previous analysis assume single SSP model and perturbed it, what
%is similar to assume a delta-function in the SFH and chemical enrichment
%distribution. In our simulations we consider a time resolved SFH and chemical enrichment, 
%what it is more realistic in our opinion.

Figure \ref{fig:Age_Met_NGC2916} shows a comparison between the input
and output coefficients of the decomposition of the stellar population
with the {\tt gsd156} template library for the spectra extracted using
a 5$\arcsec$ diameter aperture on the real and simulated CALIFA
data cube. Both distributions have a very similar pattern, with the
oldest stellar population being metal poor (sub-solar metallicity),
and a rapid chemical enrichment in the early epochs of the
universe. The star formation history presents several bursts, 
with most of the stars formed in the first 1-2 Gyr of the Universe. 
This pattern is very similar to the one
described by \citet{eperez13}.  The younger stellar populations ($<$1
Gyr) are more metal poor, which could be interpreted as a result of 
infall of more pristine gas into the inner regions, which could
explain the remanent star formation events in the very center of this
galaxy. However, it could well be that these low intensity
coefficients are just a product of the noise. In any case, the code is
able to recreate realistic SFHs and chemical enrichment patterns,
whatever their nature.

%----------------------------------------------------------------
\begin{table}[!t]
\begin{center}
\label{simtab_e_NGC2916}
\caption[Results of the simulations: emission lines in the NGC\,2916 data cube]{Results of the simulations: emission lines in the NGC\,2916 data cube.}
\begin{tabular}{rrrr}\hline\hline
\multicolumn{1}{c}{S/N} & \multicolumn{1}{c}{$\Delta$Flux/Flux} &
\multicolumn{1}{c}{$\Delta$$\sigma_v$$^1$} & \multicolumn{1}{c}{$\Delta$v$_{sys}$$^1$} 
\\\hline
30.0 &-0.01$\pm$0.10 &   0.3$\pm$11.7 &   9.9$\pm$40.6 \\
10.0 &-0.05$\pm$0.16 &  -0.8$\pm$11.7 &  11.5$\pm$32.6 \\
5.5  &-0.10$\pm$0.32 &  -3.8$\pm$15.2 &   6.1$\pm$80.1 \\
2.3  &-0.15$\pm$0.45 &   0.9$\pm$20.7 &  -0.1$\pm$99.8 \\
\hline
\hline\end{tabular}
\end{center}

(1) Kinematic parameters are given in km/s.

\end{table}
%----------------------------------------------------------------

\subsubsection{Accuracy of the  properties recovered for emission lines}
\label{simReal_elines}

Figure \ref{fig:lines_NGC2916} shows the results of the comparison
between the properties of the simulated emission lines and the
values recovered for the simulation based on the CALIFA V500 data cube
for NGC\,2916. A more quantitive summary is presented in Table
\ref{simtab_e_NGC2916}, where the average and standard deviation of
the differences between the input and output values of the different
parameters are presented for different ranges of
signal-to-noise ratio of the emission lines. Being a realistic
simulation based on a galaxy with mild emission lines, the dynamical
range for the signal-to-noise is more reduced than in the case of the
simulations presented in Sec. \ref{rEL}. The results are consistent
in general. As expected, the accuracy { and precision} in the derivation
of the different parameters is slightly lower for this 
simulation, which is more realistic in terms of the signal-to-noise
and the propagation of the imperfect subtraction of the underlying
stellar population and other sources of non-Poissonian noise. For the
well detected emission lines, S/N$>$10, the flux intensity is
recovered within 10-15\%, the velocity dispersion is recovered
within $\pm$10 km/s, and the velocity is recovered within 
30-40 km/s. The { precision} decreases for the emission lines detected with
low S/N ($<10$), and the results are unreliable for those emission lines
barely detected.

%figs/Starlight_FIT3D_age_lum_cen.ps  figs/Steckmap_FIT3D_age_lum_cen.ps  figs/Steckmap_STARLIGHT_age_lum_cen.ps
%figs/Starlight_FIT3D_met_lum_cen.ps  figs/Steckmap_FIT3D_met_lum_cen.ps  figs/Steckmap_STARLIGHT_met_lum_cen.ps

\begin{figure*}[!t]
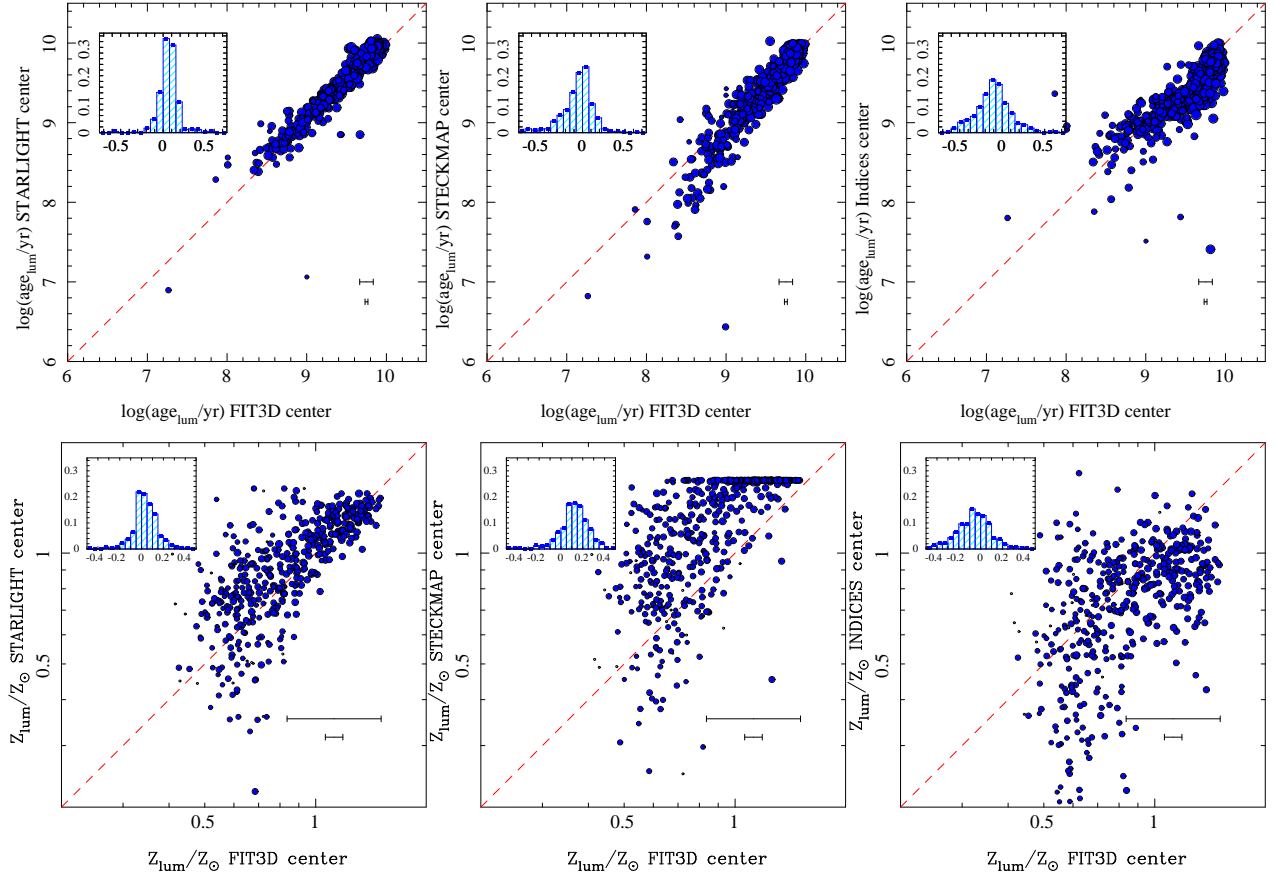

\includegraphics[angle=270,width=0.33\linewidth]{figs/Starlight_FIT3D_age_lum_cen.ps}\includegraphics[angle=270,width=0.33\linewidth]{figs/Steckmap_FIT3D_age_lum_cen.ps}\includegraphics[angle=270,width=0.33\linewidth]{figs/Indices_FIT3D_age_lum_cen.ps}
\includegraphics[angle=270,width=0.33\linewidth]{figs/Starlight_FIT3D_met_lum_cen.ps}\includegraphics[angle=270,width=0.33\linewidth]{figs/Steckmap_FIT3D_met_lum_cen.ps}\includegraphics[angle=270,width=0.33\linewidth]{figs/Indices_FIT3D_met_lum_cen.ps}
\caption{Comparison between the derived parameters of the stellar population for 448 central spectra extracted from the CALIFA V500 data cubes analyzed using {\sc FIT3D},  {\sc Starlight}, {\sc Steckmap}, and absorption line indices using the same template of stellar populations. The top panels show the comparison between the luminosity-weighted ages, and the bottom panel shows the comparison between the luminosity-weighted metallicities. In all the panels the size of the symbols are inversely proportional to the S/N of the spectrum analyzed. The error bars indicate the larger and average errors in each of the parameters, as derived by {\sc FIT3D}. In each panel the inset box shows the normalized histograms of the differences between the two parameters compared.}  \label{fig:starlight}
\end{figure*}

%\begin{figure*}[!t]
%\includegraphics[angle=270,width=0.33\linewidth]{figs/Starlight_FIT3D_age_lum.ps}\includegraphics[angle=270,width=0.33\linewidth]{figs/Steckmap_FIT3D_age_lum.ps}\includegraphics[angle=270,width=0.33\linewidth]{figs/mean_FIT3D_age_lum.ps}
%\includegraphics[angle=270,width=0.33\linewidth]{figs/Starlight_FIT3D_met_lum.ps}\includegraphics[angle=270,width=0.33\linewidth]{figs/Steckmap_FIT3D_met_lum.ps}\includegraphics[angle=270,width=0.33\linewidth]{figs/mean_FIT3D_met_lum.ps}
%  \caption{Comparison between the derived parameters of the stellar population for the CALIFA V500 datacube of NGC\,2916 when analyzed using {\sc FIT3D},  {\sc Starlight} and  {\sc Steckmap}, using the same template of stellar populations. The top panels show the comparison between the luminosity-weighted ages and the bottom panel shows the comparison between the luminosity-weighted metallicities. In all the panels the size of the symbols are inversely proportional to the S/N of the analyzed spectrum. Blue solid circles correspond to those spectra with a S/N$>$20, while open circles correspond to those spectra with a S/N lower than this value. The error bars indicate the larger and average errors in each of the considered parameters, as derived by {\sc FIT3D}.}  \label{fig:starlight}
%\end{figure*}

\section{Comparison with other fitting routines}
\label{comp}

As indicated in previous sections, {\sc FIT3D} is obviously not the
first package developed with the aim of recovering the properties of
the stellar population from optical spectra of galaxies. So far, we
have illustrated how well it recovers the main parameters defining the
stellar populations, based on different kinds of simulations. In the
current section we explore how those parameters recovered compare to
similar ones derived using other fitting tools.

We require a set of spectra with enough S/N in the
continuum to minimize the effects of noise in the comparison.
Therefore, we selected the central spectra extracted from the 448
CALIFA galaxies observed using the V500 setup up to September 2014,
co-adding the spaxels within a diameter of 5$\arcsec$ around the peak
intensity in the $V$-band. All those spectra have a S/N$>$50
per spectral pixel, as shown in \citet{sanchez12b}. The wide range of
galaxy masses and morphologies covered by CALIFA guarantees that
we explore a wide range of stellar populations with those spectra. We analyzed
them with {\sc FIT3D}, using the {\tt gsd156} template library,
and the procedures described in the previous sections.

Using the same data set and the same library we applied, in addition to
{\sc FIT3D}, (i) {\sc starlight} \citep{cid-fernandes05}, following
the prescriptions described in \citet{cid-fernandes13}, (ii) {\sc
  Steckmap} \citep{ocvrik2006}, following the prescriptions described
in \citet{patri11}, \citet{patri14a}, and \citet{patri14b}, and (iii)
the analysis based on the study of the absorption line indices
described in \citet{sanchez07b} and \citet{sanchez11}. In summary, 
after subtracting  the emission lines, we derive for each spectra 
a set of stellar indices (H$\delta$, H$\beta$, Mg$b$, Fe5270,
Fe5335, and D4000), and we estimate the ages and metallicities based
on a comparison with the expected values for each pair of indices with
the corresponding ones for the same stellar template, using the
{\sc rmodel} program \citet{Cardiel:2003p3435}. It is beyond the scope
of this article to describe in detail the different codes, however, we
should note that they rely on very different assumptions and
inversion algorithms. In essence, {\sc starlight} is more similar to
{\sc FIT3D} since it does a pseudo-random exploration of the parameter space
based on a Markov chain method. However, as indicated before {\sc
 FIT3D} does not adopt the Markov chain, and performs a MC realization
process and a linear inversion to derive the weights of the stellar
decomposition. None of these two codes makes any {\it a priori} assumption on the
shape of the star-formation history. On the contrary, {\sc Steckmap}
performs a derivation of the star formation history using a
peak-to-peak fitting algorithm \citep{ocvrik2006}, that allows to
introduce a Bayesian prior on the shape of the star formation history
\citep[e.g.][]{patri14a}. Finally, the analysis of the stellar indices 
presents larger uncertainties for a similar S/N level, as demonstrated by 
\citet{patri11}. However, it is the method that requires less number of assumptions
and computationally is the simplest: It is a derivation, not a fitting,
and it compares with a single SSP model, not with a combination of them. An additional difference
is that both the indices and {\sc Steckmap} do not take into account 
the effects of the dust attenuation, since they remove the continuum (which 
may have additional advantages if the spectrophotometric calibration has 
problems of any kind). In the case of the line indices
any dust attenuation affects the age and metallicity derivation, since
{\it a priori} it is assumed that it may affect D4000. Despite  all these differences, 
the main aim of the different methods is to provide with a realistic estimation of
the parameters that define the properties of the stellar population.

Figure \ref{fig:starlight} illustrates the results of this comparison.
It shows the derived luminosity-weighted ages and metallicities for the
different spectra as derived by {\sc FIT3D}, {\sc starlight},
{\sc Steckmap}, and the analysis of the stellar indices. The
correspondence between the luminosity-weighted ages is very good for
all, {\sc starlight} ($\Delta age=-$0.06$\pm$0.15 dex), {\sc Steckmap} 
($\Delta age=$0.03$\pm$0.22 dex), 
and the stellar indices ($\Delta age=$0.06$\pm$0.15 dex).
{ However, the distribution is compatible with a one-to-one relation taking
into account the estimated errors only for the comparison with {\sc starlight}, with
a $\chi^2/\nu=$1.14.}
The largest
discrepancies correspond to the spectra with the youngest stellar
populations, in particular for {\sc Steckmap}. These populations are
also the ones that present the strongest dust attenuations, and maybe
this is the reason for this discrepancy. In more detail, the
agreement between the ages seems to be better with {\sc starlight},
than with {\sc Steckmap}, that presents a systematic difference
towards younger stellar populations than {\sc FIT3D}.

The agreement between the metallicities is clearly better for {\sc starlight} 
($\Delta met=-$0.03$\pm$0.10 dex), than for the other two
methods, {\sc Steckmap} ($\Delta met=$0.03$\pm$0.22 dex), and stellar
indices ($\Delta met=$0.10$\pm$0.27 dex), when compared with 
{\sc FIT3D}. { Indeed, the distribution is compatible with a one-to-one relation taking
into account the estimated errors only for the comparison with {\sc starlight}, with
a $\chi^2/\nu=$1.08.} In particular, in the range of more metal rich stellar
populations, for which {\sc Steckmap} collapses the values towards the
highest metallicities in the SSP libraries. This may reflect the
similarities in the procedure between {\sc FIT3D} and {\sc starlight},
and the differences between both of them and {\sc Steckmap}.  In the
case of the stellar indices the agreement is better for the range of
more metal rich stellar populations, for which the absorptions
sensible to the metallicity are more prominent. 

%Curiously, on average,
%{\sc starlight} derives slightly older and more metal poor stellar
%populations, and {\sc Steckmap} derives slightly younger and more metal
%rich ones, reflecting somehow a different trend in the age-metallicity
%degeneracy.

As a sanity check we perform the same comparison for the values
derived between both {\sc starlight} and {\sc Steckmap}, and we find
similar results. The dispersion in the differences between the 
parameters derived is $\Delta age =$0.10$\pm$0.26 dex, and $\Delta met
=-$0.10$\pm$0.14 dex, respectively, confirming the trend described when
comparing both procedures with {\sc FIT3D}. Curiously, the values
derived with {\sc FIT3D} seem to correspond somehow to the
average/intermediate values derived by both methods. If the
discrepancies were dominated by systematic effects or differences
between the three methods, this should not be the case (appart from
the effect in the metallicity when comparing with {\sc Steckmap}).  For the
stellar indices we found that the differences in ages and
metallicities correlate with dust attenuation, as a clear
indication that this method tries to compensate the effects of dust by
assuming an older and more metal rich stellar population.

%This test highlights that not all the parameters that define the
%stellar populations are recovered with equal precision, as already
%noticed by previous studies \citep[e.g.][]{cid-fernandes14}, and in
%previous sections.  

\begin{figure}[!t]
\includegraphics[angle=270,width=1.0\linewidth]{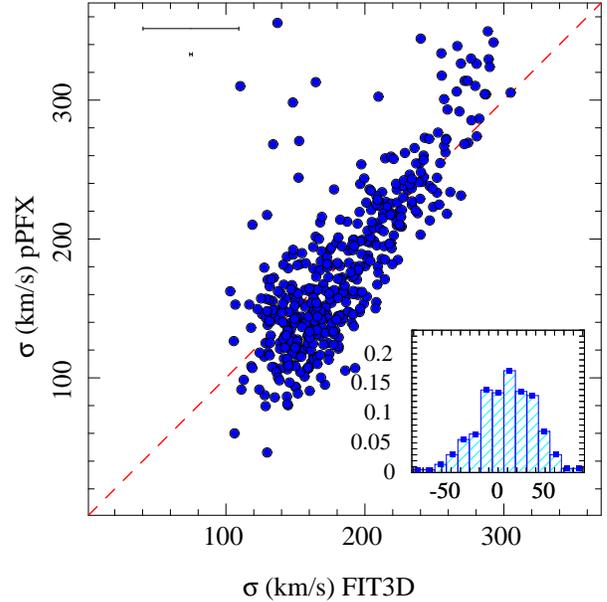}
\caption{Comparison between the stellar velocity dispersion derived for 448 central spectra extracted from the CALIFA V500 data cubes analyzed using {\sc FIT3D} and {\sc pPFX}. The inset box shows the normalized histograms of the differences between the two velocity dispersion measurements.}  \label{fig:disp}
\end{figure}

\subsection{Comparison of the stellar velocity dispersion}

{\sc FIT3D} provides  the main parameters of the stellar
populations, including the stellar kinematics. In general, the stellar
velocity is very well recovered by most of the stellar
population analysis procedures with a good accuracy { and precision}. However, the
stellar velocity dispersion is not recovered equally well, mostly due
to the degeneracy between velocity dispersion and metallicity \citep[e.g.][]{patri11}. 

It is widely accepted that one of the best procedures to analyze the
stellar kinematics is {\sc pPXF} \citep{cappellari04}. This routine
allows to recover the line-of-sight velocity distribution (LOSVD)
from absorption line spectra with high accuracy, allowing to model it
not only with a simple Gaussian, characterized by its 
velocity and velocity dispersion, but with more complex profiles for
the LOSVD, including asymmetries characterized by Hermite functions
and/or multiple Gaussian-like velocity components. This is the reason
why it  is widely used in stellar population analyses,
even when additional  tools are needed to derive the other main properties
of  those stellar populations \citep[e.g.][]{patri14a,patri14b}. The current capabilities
of {\sc pPXF} regarding the analysis of the stellar kinematics are far
beyond the  procedures implemented in the improved version of {\sc FIT3D}. 
However, it is worth comparing the parameters derived, in particular
the stellar velocity dispersion, to illustrate the precision of our current
estimations.

Figure \ref{fig:disp} shows the comparison between the stellar
velocity dispersion derived using {\sc FIT3D} with the one derived
using {\sc pPXF} for the set of spectra described in
Sec. \ref{comp}. There is a very good agreement between the
estimations done by both procedures ($\Delta\sigma=-$9.8$\pm$36.7
km/s), with a difference compatible with the expected { precision} for the
given spectral resolution 
\citep[$R\sim$850][]{sanchez12b}. Therefore, despite the fact that
the current version of {\sc FIT3D} is not optimized for the analysis of
the kinematic properties of the stellar populations, it provides
an accurate estimation of stellar velocity dispersion.

In summary, we can conclude that parameters of the
stellar populations derived with {\sc FIT3D} are consistent with those
provided by other algorithms, with the difference that this
new algorithm provides a reliable estimation of the errors of
those parameters.

\section{Summary and conclusions}
\label{summ}

In this article we present a new version of {\sc FIT3D}, a package developed 
to analyze the optical spectra of galaxies, in order to derive the
main properties of their stellar population and the ionized gas.  {\sc
  FIT3D} implements a new MC iteration method that provides not only
the numerical values of the parameters, but also an estimation of their errors. 
In this regard the procedure is a substantial modification
of the previous version of the package.

The main results of this study are the following:

\begin{itemize}

\item The fitting routine is confronted with simulations that
  demonstrate its ability to recover the properties of the underlying
  stellar population and the emission lines. We found that the
  algorithm is able to recover the luminosity weighted ages and
  metallicities with a { precision} of $\sim$0.2 dex for spectra
  with a S/N per pixel better than $\sim$50, for most of the stellar
  templates analyzed in the simulations. We { found} systematic
  differences/offsets in the recovered parameters. { However, they
  are sub-dominant for the individual realizations, and only significant
  in an statistical way. We recover the well } known age - metallicity - dust attenuation degeneracies.  In
  general, the degeneracies are of the order of the precision in the
  parameters recovered. The use of different templates with different
  spectral resolutions and coverage of the parameter space affects the
  accuracy and { precision} of the parameters recovered in a similar way as a decrease
  of the S/N.  In this regard, it is recommended to use simulations
  {\it ad hoc} in order to determine which stellar templates are the
  optimal for each particular data set. In general we cannot recover
  the luminosity-weighted age, metallicity, and dust attenuation with
  a precision better than $\sim$0.1 dex, even for the highest
  signal-to-noise.

\item The algorithm is not tuned to derive an accurate/precise
  stellar kinematics. However, it provides  a good estimation of
  the stellar velocity and velocity dispersion, that at a typical
  spectral resolution of $R\sim$1000 is never better than $\sim$25
  km/s, for the highest S/N explored. The current implemented
  approach to derive the kinematics introduces a systematic effect in the
  profile of the  absorption lines subtracted. This is translated to an unrealistic
  emission that is at most of the order of EW(H$\alpha$)$\sim$0.4\AA, 
  far below the typical detection limit for most of spectroscopic surveys \citep[e.g., Gomes et al., in prep.,][]{papa13}.

\item The program is able to recover the star formation and chemical enrichment histories
for the galaxies when the signal-to-noise per pixel is large enough (S/N$>$50).
The precision in the ages and metallicities recovered is better when the SFH comprises
correlations between the distribution of the stellar populations than in the case of
a simple distortion of a single burst of fixed age and metallicity. 

\item The simulations indicate that the nominal errors derived by {\sc
  FIT3D} for the parameters that characterize the stellar populations
  are a factor four smaller than the estimated errors derived from the
  comparison of the input and estimated values for the stellar ages,
  metallicities, and dust attenuation.

\item A detailed set of simulations allow us to determine the ability of
  the algorithm to recover the intensity, velocity dispersion, and
  velocity of the emission lines in three main cases: isolated lines,
  blended lines, and blended lines with an underlying broad emission
  line component. In general, it is possible to estimate the 
  parameters with a precision of 0.05-0.10 dex for S/N above 3$\sigma$. 
  As expected, the { precision} of the  parameters derived
  is worse for the most blended emission lines and for the lower S/N.

\item A more realistic set of simulations was created based on real
  spectra of a typical galaxy extracted from the CALIFA survey
  data cubes, by using as input for the simulations the output of the
  analysis of this galaxy, based on the same algorithm, and
  perturbed by a realistic realization of the noise, based on the
  residuals of the analysis described. These simulations confirm that
  we can recover the properties of the stellar populations with a
  precision of the order of $\sim$0.1 dex only for S/N$\sim$50.

\item The stellar parameters derived using {\sc FIT3D} are compared with similar
ones derived using widely used fitting routines, like {\sc starlight} and {\sc Steckmap},
and with those derived using classical stellar indices, for a set of
high S/N spectra extracted from the CALIFA data cubes. This comparison shows that
the parameters derived by {\sc FIT3D} are compatible with those  derived 
using the other  tools. In particular, there is a very good agreement in the luminosity-weighted 
ages for the three cases, with a typical difference of  $\sim$0.15-0.22 dex.
The agreement is slightly worse for the luminosity-weighted metallicities in the case
of {\sc Steckmap} and the stellar indices, being better for the case of {\sc starlight}, with
a typical difference of  $\sim$0.1 dex.

\end{itemize}

In summary, we demontrate that the new algorithm implemented in {\sc
  FIT3D} is able to recover the main properties of the stellar
populations and emission lines for moderate resolution spectra of
galaxies. The tool is fully prepared to analyze IFS data, and in
particular is well suited for the analysis of IFU data from the
major on-going surveys, like CALIFA \citep{sanchez12b}, MaNGA
\citep{manga}, and SAMI \citep{sami}. In forthcoming articles we will
present a practical implementation to analyze spatially resolved
spectroscopic data and a study of the possible uncertainties.

\begin{acknowledgements}

We acknowledge the referee for his/her comments and suggestions that has
helped to improve the manuscript.

SFS thanks the director of CEFCA, M. Moles, for his sincere support to
this project.

We thanks C.J. Walcher and R. Gonz\'alez Delgador for their valuables
comments and suggestions that has improved this manuscripts in many
senses.

SFS thanks the CONACYT-125180 and DGAPA-IA100815 projects for providing him support in this
study.

We acknowledge support from the Spanish Ministerio de Econom\'\i a y Competitividad, 
through projects AYA2010-15081 and AYA2010-10904E, and Junta de Andaluc\'\i a FQ1580.
EP acknowledges support from the IA-UNAM and from the Guillermo Haro program at INAOE.

This study  uses data provided by the Calar Alto Legacy
Integral Field Area (CALIFA) survey (http://califa.caha.es/).

CALIFA is the first legacy survey performed at Calar Alto. The
CALIFA collaboration would like to thank the IAA-CSIC and MPIA-MPG as
major partners of the observatory, and CAHA itself, for the unique
access to telescope time and support in manpower and infrastructures.
The CALIFA collaboration also thanks the CAHA staff for the dedication
to this project.

Based on observations collected at the Centro Astron\'omico Hispano
Alem\'an (CAHA) at Calar Alto, operated jointly by the
Max-Planck-Institut f\"ur Astronomie and the Instituto de Astrof\'\i sica de
Andaluc\'\i a (CSIC).

\end{acknowledgements}

\bibliography{CALIFAI}
\bibliographystyle{mdpi}

%\]
\end{document}